\documentclass[aps,prd,twocolumn,showpacs,superscriptaddress,floatfix,showkeys,nofootinbib]{revtex4-2}
\usepackage[english]{babel}
\usepackage[utf8]{inputenc}
\usepackage[colorinlistoftodos, color=green!40, prependcaption]{todonotes}
\usepackage{amsthm}
\usepackage{mathtools}
\usepackage{physics}
\usepackage{xcolor}
\usepackage{graphicx}
\usepackage[left=23mm,right=13mm,top=35mm,columnsep=15pt]{geometry} 
\usepackage{adjustbox}
\usepackage{placeins}
\usepackage[T1]{fontenc}
\usepackage{lipsum}
\usepackage{csquotes}
\usepackage[pdftex, pdftitle={Article}, pdfauthor={Author}]{hyperref} % For hyperlinks in the PDF
\usepackage{bookmark}
\usepackage{booktabs}
\usepackage{amsmath}
\usepackage{orcidlink}

\usepackage[commentmarkup=uwave,deletedmarkup=xout]{changes}
\definechangesauthor[name={Nicole}, color=purple]{NK}
\definechangesauthor[name={Will}, color=orange]{WF}
\definechangesauthor[name={Max}, color=violet]{MI}

\newcommand\pasa{\ref@jnl{PASA}}%  % Publications of the Astron. Soc. of Australia
\newcommand{\ccaaffil}{\affiliation{Center for Computational Astrophysics, Flatiron Institute, New York NY 10010, USA}}
\newcommand{\cuaffil}{\affiliation{
    Columbia University, Department of Astronomy, 550 West 120th Street, New York, NY, 10027, USA
}}
\newcommand{\sbuaffil}{\affiliation{Department of Physics and Astronomy, Stony Brook University, Stony Brook NY 11794, USA}}

\begin{document}
\title{Polarization Analysis of Ringdown Signals}

\author{Nicole Khusid\,\orcidlink{0000-0001-9304-7075}}
    \email[Contact me: ]{nicole.khusid@stonybrook.edu}
    \sbuaffil
    \ccaaffil

\author{Will M. Farr\,\orcidlink{0000-0003-1540-8562}}
    \sbuaffil
    \ccaaffil

\author{Maximiliano Isi\,\orcidlink{0000-0001-8830-8672}}
    \cuaffil
    \ccaaffil

\date{\today} % Leave empty to omit a date

\begin{abstract}
Merging binary black holes exhibit a ringdown phase in which they primarily emit gravitational waves in the shape of damped sinusoids corresponding to quasi-normal modes of the Kerr remnant. In general, each mode carries four degrees of freedom encoding amplitude and phase information. When the modes are excited with equatorial reflection symmetry, as is the case for black hole mergers with spins (anti)aligned to the orbital angular momentum, the symmetry constrains two degrees of freedom. As a result, the relationship between 
% \mi{relative to what? this sounds like you mean the amplitudes of each mode relative to the others, which is not right} 
polarization amplitudes and phases in each mode is fixed by the viewing (inclination) angle to the equatorial plane. We use such a constrained model to fit the ringdown signals of both non-precessing and precessing systems such as GW150914 and GW190521, respectively. We show that we can measure the degree of circular polarization and handedness
% \mi{not super transparent, maybe say: ``the degree of circular polarization''} 
of ringdown signals like those of GW150914, even if only the two LIGO detectors are available; such a polarization measurement can be translated into an inferred source inclination assuming the reflection symmetry above, again using the ringdown signal alone.
    % For precessing systems with spin-orbit misalignment, mismodeling the polarization structure of their ringdowns induces a bias in the measurements of remnant BH properties and mode parameters of such signals.
On the other hand, the constrained polarization model is insufficient to capture the polarization structure of signals from precessing systems, leading to biases in the inferred mode frequencies and amplitudes.
We explore the magnitude of this effect by fitting GW190521-like injections with the restricted model, finding weaker predictive accuracy relative to the arbitrary-polarization model and potentially significant systematic biases.
% \mi{this is not a very effective way of presenting your result! Say first that we find that mismodeling the polarization does induce a bias, then caveat that it will be mostly relevant for higher sensitivity} 
% Despite the \replaced[id=NK]{constrained polarization model lacking the physics necessary to robustly infer the}{model's apparent limitations in accurately recovering injected} remnant BH properties and mode parameters \added[id=NK]{of such signals}, we find its predictive accuracy is comparable to that of a free polarization model for present-day and doubly improved instrument sensitivity. 
As our detectors continue to improve, using the correct polarization model is increasingly important to avoid biased ringdown measurements.
\end{abstract}


\maketitle

\section{Introduction} \label{sec:intro}
% Gravitational waves (GWs) arise from the mergers of compact objects throughout the Universe. Those GWs arising from the mergers of stellar mass sources can be observed by ground-based detectors that make up the LIGO-Virgo-KAGRA (LVK) network \cite{2015CQGra..32g4001L, 2015CQGra..32b4001A, 2021PTEP.2021eA101A}. The large majority of mergers detected by the LVK are consistent with quasi-circular binary black hole (BBH) systems \cite{2019PhRvX...9c1040A,2021PhRvX..11b1053A,2023PhRvX..13d1039A}. Upon merger, the GW signals from these sources are quickly dominated by the ringdown phase. Ringdown, in the theory of general relativity (GR), is the process by which a perturbed BH radiates GWs down to a stable, Kerr BH \cite{1963PhRvL..11..237K,1965JMP.....6..915N,1973ApJ...185..635T,1973ApJ...185..649P,1974ApJ...193..443T,1999JApA...20..269A,2015CQGra..32l4006T,2021PhRvD.103j4017L}. Soon after merger, the GW radiation emitted throughout this process is comprised of linearly superposed damped sinusoidal quasi-normal modes (QNMs) whose frequencies and damping rates are determined by the final mass and spin of the remnant BH according to the No Hair Theorem; the practice of measuring the QNM spectrum in a ringdown signal to infer remnant BH properties is called BH spectroscopy.

Gravitational waves (GWs) are emitted throughout the coalescence of binary black holes (BBHs) \citep{Blanchet2014, GW150914}. In the ringdown phase, during which the remnant settles down to a stable Kerr black hole (BH), the gravitational radiation is described by perturbation theory \citep{1963PhRvL..11..237K,Vishveshwara1970b,1973ApJ...185..635T,1973ApJ...185..649P,1974ApJ...193..443T,BertiRingdownReview2025}. Soon after the merger, the GW signal is well modeled as a superposition of elliptically polarized modes indexed by (azimuthal, magnetic, overtone) numbers $(\ell, |m|, n)$, comprised of pairs of contributions from right- and left-handed ($+|m|$ and $-|m|$, respectively) circularly polarized damped sinusoids called quasinormal modes (QNMs) \citep{Vishveshwara1970b, Press1971,ChandrasekharDetweiler1975,Leaver1985,Berti+2009,London+2014,2019PhRvL.123k1102I,Giesler+2019,Giesler+2025,BertiRingdownReview2025}. By the no-hair theorem in general relativity, the frequencies and damping rates of these modes are solely determined by the mass and spin of the remnant BH, permitting spectroscopic measurements of remnant properties through the inference of QNM characteristics \citep{Carter,Penrose:1969pc,Detweiler:1980gk,Dreyer:2003bv,Berti+2009,Gurlebeck:2015xpa,BertiRingdownReview2025}. 

In the most general situation, every $|m|$ mode is comprised of two polarization states---plus ($+$) and cross ($\times$)---that each have a free amplitude and phase determined by the initial perturbation and the source viewing angle \cite{2021arXiv210705609I,Isi2023}, so every mode carries four polarization degrees of freedom. 
% To visualize this generic polarization structure, consider a coordinate system called the ``wave frame" that is aligned with the GW itself, so one axis lies along the wave propagation direction and the other two axes are transverse to it \citep{Isi2023}. We can define the $+$ and $\times$ polarization axes in the linear basis within the transverse plane of the wave frame. In this plane, the decomposed strain for each mode, $h_{\ell |m| n}$, can be represented as a vector whose tip traces out a curve over time called the ``polarization ellipse'' (see Fig. 3 of~\cite{Isi2023}).
These amplitudes and phases encode the ringdown mode's relative amounts of circular and linear polarization, which may 
% The shape of each mode's polarization ellipse, or the \textit{ellipticity} of the mode, encodes the relative amplitudes and phases of its $+$ and $\times$ polarization states and can 
evolve with time based on the dynamics of the source \citep{2021arXiv210705609I,Isi2023}.
% \mi{this is compressing way too much info and introducing too many concepts too quickly---do you need to introduce all of these new terms (wave frame, polarization ellipse, ellipticity, linear basis)? and if so, does it all have to happen here? at this stage I think you can just say that we can characterize each mode by its degree of circular vs linear polarization. better to introduce the other concepts in the next section as needed}
% In the most general situation, each mode carries four quadrature \mi{``quadrature'' is jargon, and totally not transparent to people other than us. Better to first introduce the notion of there being two polarization states per mode, each with an amplitude and a phase. In fact, the concept of ``quadrature'' is totally unecessary at this stage, and probably at any stage, except the technical discussion of how our likelihood is written down, which is ancillary to the physics} degrees of freedom---an amplitude and a phase for each of the $+m$ and $-m$ components---which are determined by the initial perturbation \cite{2021arXiv210705609I,Isi2023}.

% {Most generically, the postmerger signal can be modeled without any assumptions about these initial conditions, such that the amplitudes scaling the $\pm m$ QNM contributions to each ringdown mode, and thus the signal's polarization structure, are free.}
Symmetries of progenitor BBH systems prior to merger are expected to carry over to the excitations of QNMs on their remnants \citep{Kamaretsos+2012,London+2014,Zhu+2025}.
Most BBH mergers detected by the LIGO-Virgo-KAGRA (LVK~\cite{2015CQGra..32g4001L, 2015CQGra..32b4001A,2021PTEP.2021eA101A}) network are consistent with not measurably precessing systems whose component spins are consistent with being (anti-)aligned with the systems' orbital angular momentum \citep{GWTC4}. 
Such configurations with negligible in-plane spins give rise to symmetric source configurations and gravitational emission under a parity operation consisting of reflection about the orbital plane with axial spin reversal \citep{Boyle+2008, Boyle+2014}. This well-established equatorial reflection symmetry is expected to
% \mi{(1) what does ``invariant spacetimes'' mean? this is a very complicated way of saying that the source and radiation are symmetric under reflection about the orbital plane.}
translate into the ringdown, constraining the polarization states of each mode to share a common amplitude and phase, thus reducing to two degrees of freedom. This symmetry enforces that the polarization state of each ringdown mode is stationary and simultaneously controlled by the inclination angle, $\iota$, between the line of sight to the observer and the remnant BH's spin. In other words, for non-precessing systems, the only parameter that determines the polarization structure of the ringdown signal is $\iota$.
% \added[id=NK]{ placing physical constraints on the mode quadratures. 
% As a result, each $\pm |m|$ mode carries equal magnitude and 90$^{\circ}$ out-of-phase contributions from both GW polarizations \citep{Isi2023}. While these modes are intrinsically ``circular,'' } \mi{what does ``intrinsic'' mean in this sense? what does ``circular'' mean?} 
    % their \textit{observed} polarization structure is modulated by the inclination angle between the line of sight to the observer and the remnant's spin. 
By incorporating these constraints from equatorial reflection symmetry into a ringdown model, one may infer the inclination of a BBH system by leveraging the polarization information in its ringdown alone.
% Such a model is a subset of the most generic ringdown model, and moving forward we refer to it as the ``aligned-spin'' ringdown model.  
% \mi{The exposition here is not very effective. It kind of half assumes the reader already shares your jargon and knows what you are saying: those readers are skipping this paragrap, don't focus on them. Better to take a step back and say a few words about the polarization structure of a GW mode and how it relates to inclination if the spins are aligned.}

While previous works have employed equatorial reflection symmetry constraints to model ringdown signals from non-precessing systems, these implementations have not consistently modeled polarization degrees of freedom; some ringdown-only analyses introduce the symmetry at the source level either by constraining the intrinsic $\pm |m|$ QNM amplitudes \citep{Carullo+2019,pyRing,Gennari+2024} or by parameterizing deviations away from the symmetry \citep{Capano+2023,Capano+2024,Ghosh+2025}, while others place constraints on the detector-frame waveform by enforcing a fixed relationship between observed polarizations \citep{2019PhRvL.123k1102I,2021arXiv210705609I}. Moreover, waveform templates that model full GW signals have different descriptions of the ringdown portion \citep{IMRPhenom,SEOBNRv5}, so the implicitly encoded polarization structure in these templates also varies.
Especially for ringdown-dominated signals, inconsistent polarization prescriptions between full-signal and ringdown-only analyses could result in discrepant inference.
% These inconsistencies could result in biased parameter estimation, as well as disagreement with ringdown-only inference.
% \mi{which inconsistencies? between these different models? why would that cause biases in parameter estimation? Or do you mean that individual models aren't self-consistent?}
% \mi{It would be better to motivate this paper by saying that the modeling of the polarization DOFs has been inconsistent in the literature: sometimes restrictive, sometimes free (and provide citations). Offer this paper as a solution to clear up that mess.}

In this paper, we demonstrate the importance of carefully modeling the polarization degrees of freedom in ringdown analyses to extract the most information from the signal and avoid systematic biases in the inference of remnant properties.
% We employ a ringdown model that explicitly ties the polarization structure of the signal to a measurement of the system's inclination angle.
% In this work, we extend these efforts with \deleted[id=NK]{an aligned-spin}{a} ringdown model that ties the polarization structure of the signal to a measurement of the system's inclination angle \mi{that is not the main highlight of this paper: such models were the first type of models considered (even if in a more restricted form). the main highlight is what you find, and how you show that modeling this correctly matters.}. 
Measuring polarization content from the ringdown alone can be difficult, as the signal-to-noise ratio (SNR) of the truncated signal is often much more modest compared to that of the full inspiral-merger-ringdown (IMR) signal. Despite the SNR decrease, ringdown analyses are compelling for their simplicity, which may bypass potential waveform systematics arising from full-signal (i.e., IMR) analyses.

We extend the most generic ringdown model to leverage the assumptions of equatorial reflection symmetry by explicitly tying signal polarization structure to inclination angle measurements and hereafter call this the ``aligned-spin'' model. The aligned-spin model is implemented within the \textsc{ringdown} software package \citep{2021arXiv210705609I,ringdown}, which we use to produce all results presented in this paper. We compare the performance of the aligned-spin model to that of the generic, unconstrained model in fitting real and synthetic ringdown signals from both non-precessing and precessing systems. As proxies for such systems, we use GW150914 and GW190521, whose IMR-inferred parameters are consistent with obeying and breaking equatorial reflection symmetry, respectively. We show that, for SNRs that are already achievable in the most recent LVK observing run \cite{GW230814,AreaLaw}, mismodeling the polarization information present in ringdown signals results in significantly biased measurements of remnant properties. This bias can serve as evidence of precession from the ringdown alone, independently of conclusions from IMR analyses.

The potential for biased inference in loud observations is an especially important concern for heavy systems, like GW231123, whose GW signals are dominated by the ringdown and benefit from ringdown-only inference to corroborate IMR analyses in poorly understood regions of parameter space \citep{GW231123,Siegel+2025}. As the sensitivity of the detector network improves and the detected GWs grow in SNR, our signal templates must take care to properly model polarization structure to produce reliable parameter estimation. We show in this work that ringdown-only analyses can extract this information as a complementary polarization measurement or, importantly, as a litmus test of systematics due to mismodeling.

This paper is organized as follows. In Section \ref{sec:ii} we introduce the formalism of the aligned-spin ringdown model. In Section \ref{sec:iii} we compare analyses of GW150914's ringdown using both the aligned-spin and generic models. Finally, in Section \ref{sec:iv} we investigate the validity of the aligned-spin model in cases where it is \textit{not} physically relevant, i.e. synthetic signals from precessing binaries similar to GW190521. The Appendix contains useful technical details regarding the implementation of the aligned-spin model.
% BBHs with individual spins that are perpendicular to the system's
% orbital angular momentum, be they aligned or anti-aligned, give
% rise to remnant BHs whose spins are still perpendicular to the
% equatorial plane. In modeling the ringdown signal for such a 
% system, the symmetry requires geometric constraints  
\section{The Aligned-Spin Formalism} \label{sec:ii}
% \subsection{Formalism} \label{subsec:iiA}

To excellent approximation, the total ringdown emission, $h = h^{+} - ih^{\times}$, at sufficiently late times decomposes into a linear superposition of damped sinusoidal QNMs, projected onto spin-weighted spheroidal harmonic basis functions that solve the angular portion of the linear perturbation equations for Kerr BHs \citep{1973ApJ...185..635T,1973ApJ...185..649P,Leaver1985,Berti+2006,Cook&Zalutskiy}.
Because waveform templates employ a spin-weighted spherical harmonic decomposition instead, spheroidal-spherical mode mixing arises as a result of this basis mismatch in the \textit{modeling} of the radiation on the sky. For modes like the (2,2,0) that typically dominate ringdown signals, the mixing is negligible, such that both bases are approximately interchangeable \citep{KellyBaker2013,BertiKlein2014}. We proceed with the spin-weighted spherical harmonics, $_{-2}Y_{\ell m}$.
% \mi{This is a round-about way of saying we will approximate spheroidal harmonics as spherical harmonics and that we incur some small error in doing so. The paragraph introduces many conceprs that are not needed and get in the way of the main point.}
Summing contributions from pairs of $\pm m$ QNMs, we can write down an expression for each QNM as a purely polarized (elliptical) mode $h_{\ell|m|n}$; namely, following \cite{2021arXiv210705609I},
\begin{align} \label{eq:qnm_gen_exp}
    h_{\ell|m|n}(t) = &\left[ C_{\ell mn}e^{-i\omega_{\ell|m|n}t}\, _{-2}Y_{\ell m}(\iota, \varphi) \right. + \\
    &\left.\phantom{[} C_{\ell -mn}e^{i\omega_{\ell|m|n}t}\, _{-2}Y_{\ell -m}(\iota, \varphi) \right] e^{-t/ \tau_{\ell|m|n}} \, , \nonumber
\end{align}
where $C_{\ell mn}$ are the intrinsic complex QNM amplitudes, $\omega_{\ell|m|n}$ and $\tau_{\ell|m|n}$ are mode frequencies and damping times, respectively, and the spin-weighted spherical harmonics $_{-2} Y_{\ell m}$ are functions of the inclination between the line of sight to the GW source and the orbital angular momentum vector, $\iota$, and azimuthal angle $\varphi$.

For BBH systems that obey equatorial plane symmetry, the component spins are either aligned or anti-aligned with the orbital angular momentum. Once the binary merges, the remnant preserves this symmetry and retains a final spin that is aligned or anti-aligned with the equatorial plane normal vector. Perturbations on the final BH during the ringdown phase inherit this symmetry \citep{1973ApJ...185..635T,Berti+2009,Kamaretsos+2012,London+2014,Zhu+2025} such that the GW emission is complex-conjugated under reflections about the equatorial plane ($\iota \rightarrow \pi - \iota$), implying that the intrinsic amplitudes of contributing $\pm |m|$ QNMs obey $C_{\ell -mn} = (-1)^{\ell}  C^*_{\ell mn}$ (see Sec. IIIC of \cite{Isi2023}). 

With the above parity constraint, and taking $C_{\ell mn} = \bigl|C_{\ell mn}\bigr|e^{i\phi_{\ell mn}} \equiv \mathcal{A}_{\ell m n}e^{i\phi_{\ell mn}}$, we can simplify Eq.~\eqref{eq:qnm_gen_exp} to
\begin{align} \label{eq:qnm_ali}
    &h_{\ell|m|n}(t) =  \\
    &\phantom{-\, i\,}\left[_{-2}Y_{\ell|m|}(\iota) + _{-2}Y_{\ell|m|}(\pi-\iota) \right] \mathcal{A}_{\ell|m|n}(t) \cos\Phi_{\ell|m|n}(t) \nonumber \\
     &-i \left[_{-2}Y_{\ell|m|}(\iota) - _{-2}Y_{\ell|m|}(\pi-\iota) \right] \mathcal{A}_{\ell|m|n}(t) \sin\Phi_{\ell|m|n}(t) \, , \nonumber
\end{align}
where we have used the property $_{-2}Y_{\ell -m}(\iota, \varphi) = (-1)^{\ell}  _{-2}Y^{\ast}_{\ell m}(\pi-\iota, \varphi)$ \citep{SWSH,1973ApJ...185..649P,Cook&Zalutskiy}, extracted the $\phi$ and $\varphi$--dependence into the overall phase $\Phi_{\ell|m|n}(t)$, let $\mathcal{A}_{\ell m n}(t) = \mathcal{A}_{\ell m n}e^{-t / \tau_{\ell |m| n}}$, and expanded complex exponentials into sinusoids. Recalling that $h = h^{+} - ih^{\times}$, we define
\begin{equation}
    Y^{+/\times}_{\ell|m|}(\iota) = \,_{-2}Y_{\ell|m|}(\iota) \, \pm \, _{-2}Y_{\ell|m|}(\pi-\iota)
    \label{eq:ang_facs}
\end{equation}
such that the real and imaginary parts of Eq.~\eqref{eq:qnm_ali} reduce to the two polarization states (with exponential decay restored)
\begin{subequations}
\begin{align}
    h_{\ell|m|n}^+(t) = \mathcal{A}_{\ell|m|n}\,Y^+_{\ell|m|}(\iota)\cos{\Phi_{\ell|m|n}(t)} e^{-t/\tau_{\ell|m|n}} \,,\, \\ h_{\ell|m|n}^{\times}(t) = \mathcal{A}_{\ell|m|n}\,Y^{\times}_{\ell|m|}(\iota)\sin{\Phi_{\ell|m|n}(t)} e^{-t/\tau_{\ell|m|n}} \,.
\end{align}
    \label{eq:ali_pols}
\end{subequations}
If the system is face-on ($\iota = 0$) or face-off ($\iota = \pi$), such that the angular factors have equal magnitudes, each mode manifests equal contributions from both polarizations with a fixed $90^{\circ}$ phase offset, so that the wave is circularly polarized. For any other orientation, the projection of the polarization states into the plane perpendicular to the line of sight to the GW source modulates their relative amplitudes through the inclination-dependent $Y^{(+/\times)}$ in Eq.~\eqref{eq:ali_pols}, stretching the curve traced out by each mode's strain vector in the $h^+$--~$h^{\times}$ plane (see Sec. \ref{sec:intro}) from a circle into an ellipse. The mixture of $+$ and $\times$ for each mode in the ringdown signal, or its ellipticity $\epsilon_{\ell|m|n}$ \citep{2021arXiv210705609I,Isi2023}, thus depends only on the inclination of the source:
% From this expression, it is evident \mi{why is it evident? don't assume the reader is in your head! You can just state: ``this corresponds to a circularly polarized mode other than for the inclination-dependent angular factors...'', and cite, e.g., \cite{Isi2023} for more details} that each ringdown mode would be circularly polarized if not for the inclination-dependent angular factors in Eq.~\eqref{eq:ang_facs} (unless the system is face-on/off, the projection of the polarization states into the plane perpendicular to the line of sight to the GW source introduces an effective ellipticity; this is more apparent in terms of the polarization ellipse in Fig. 3 of \cite{Isi2023} for elliptically polarized waves) \mi{this parenthetical note should just be incorporated into the main text; or you can make it a footnote, depending on how you rephrase}.
\begin{equation}
    \epsilon_{l |m| n}(\iota) = \frac{\big|\,_{-2}Y_{\ell m}(\iota)\,\big| - \big|\,_{-2}Y_{\ell -m}(\iota)\,\big|}{\big|\,_{-2}Y_{\ell m}(\iota)\,\big| + \big|\,_{-2}Y_{\ell -m}(\iota)\,\big|} \, .
    \label{eq:ellip}
\end{equation}
We construct the aligned-spin ringdown model by enforcing the above constraints on polarization amplitudes and phases.
% \added[id=NK]{The polarization structure of the ringdown emission from a quasi-circular non-precessing system can be directly derived from a measurement of its inclination, as a consequence of reducing from four to two free polarization quadratures.} 
% \mi{concept of ``quadratures'' is unnecessary at this stage}

Having worked out the implications of equatorial reflection symmetry for the polarization structure of ringdown signals, we define two quantities to parameterize the particular constraints that reduce the polarization degrees of freedom from the generic model to the aligned-spin model. One measures how well correlated the measured mode ellipticities are with a global inclination parameter as in Eq.~\eqref{eq:ellip}; the other measures how well the intrinsic complex $\pm |m|$ QNM amplitudes obey the complex-conjugate relationship under equatorial reflections. To define the former, we invert Eq. \ref{eq:ellip} to compute the induced $\cos\iota$ from each $\epsilon_{\ell|m|n}$ and calculate the pairwise differences between modes:
\begin{equation}
    \delta x_k = \frac{1}{2}|\Delta\cos\iota|_k\, ,
    \label{eq:dx_main}
\end{equation}
where $1/2$ is a normalization, and $k$ is the particular pair of modes for which the difference in $\cos\iota$ is computed.
We define the latter as $y_{\ell |m| n}$,
\begin{equation}
    y_{\ell|m|n} = \frac{\bigl|C_{\ell-|m|n} - (-1)^{\ell}C^{*}_{\ell|m|n}\bigr|}{\bigl|C_{\ell|m|n}\bigr| + \bigl|C_{\ell-|m|n}\bigr|}\, ,
    \label{eq:y_intro}
\end{equation}
where the denominator is a normalization factor that ensures $y_{\ell |m| n} \in [0, 1]$, and the numerator quantifies the deviation from the intrinsic mode amplitude relationship under equatorial reflections.
% The aligned-spin ringdown model implements these reflection symmetry constraints, effectively reducing the number of free parameters per mode (when there is more than one mode in the model). 

In the free, or generic, polarization model, each elliptically polarized ringdown mode is parameterized by $\{\omega, \tau, A, \phi, \epsilon, \theta\}_{\ell|m|n}$: the mode angular frequency, damping time, amplitude, phase, ellipticity, angle defining orientation of polarization ellipse (see Fig. 3 of \cite{Isi2023}). The aligned-spin ringdown model imposes the constraints $\delta x_k = 0$ per Eq. \eqref{eq:ellip} and $y_{\ell |m| n} = 0$ and additionally samples a global polarization angle $\Delta\psi$, which replaces the per-mode polarization angle $\theta_{\ell |m| n}$.
This restricted model defines a subspace of the generic model.

With the reparameterization of $\epsilon_{\ell |m| n}$ as a function of $\iota$ in the aligned-spin model comes a change in prior: the Jacobian of the transformation in Eq.~\eqref{eq:ellip} between $\epsilon_{\ell |m| n}$ and $\cos\iota$ induces divergences (zero-densities) at the edges of the domain when the distribution for the latter (former) is uniform \citep{Isi2023}. To robustly explore the parameter space, we implement flat priors on $\cos\iota$ and $\epsilon_{\ell |m| n}$ in the aligned-spin and generic ringdown models, respectively, resulting in the prior difference shown in the left panel of Fig. 17 in \cite{Isi2023}.
% \mi{it's okay to point to the appendix for details but all symbols should be defined and the main text should be self-contained; currently this is not the case.}

Due to the reduced number of free parameters, we expect the aligned-spin model to generally produce tighter posteriors than the generic model. See App.~\ref{subsec:app1} for more technical details of the aligned-spin model, particularly polarization degrees of freedom.

% \mi{not a dealbreaker but I don't think "aligned model" is the best name; how about z-parity model?}
\section{Non-precessing systems} \label{sec:iii}

% \mi{I don't think this is the right angle. We should pick up from the discussion in Harrison's paper about the polarization degrees of freedom. Explicitly say that the restricted model is contained within the generic model, but then point out that it there is an advantage to using the restricted model when applicable (is there? you need to argue/show this).}
While the aligned-spin ringdown model is entirely nested within the generic model, using it in physically relevant cases can be advantageous to extract the most amount of information from ringdown signals. Due to fewer degrees of freedom, we expect the
aligned-spin ringdown model to perform as well as the generic unconstrained model for signals from non-precessing systems. Furthermore, for such systems the aligned-spin model can exploit polarization measurements to infer inclination angles that the generic model cannot. To demonstrate the capability of the aligned-spin ringdown model, we analyze the ringdown signal of a system that obeys the equatorial reflection symmetry of the model. Later in Sec. \ref{subsec:ivC}, we also analyze a set of similar synthetic signals of varying SNRs to demonstrate the benefit of using the aligned-spin model when it is applicable.
    
For this test case, we choose the first ever GW detection, GW150914 \citep{2016PhRvL.116f1102A,2016PhRvL.116x1102A}. This event is consistent with a BBH merger whose components are not spinning or have small spins aligned with the orbital angular momentum. From the IMR analysis, the measured inclination of the system is most likely face-off ($\cos\iota \simeq -1$), and the polarization structure of the gravitational radiation should therefore be well-constrained to be left-handed circular. 

\subsection{Fitting real GW150914 data} \label{subsec:iii_fitting}

Previous work on the ringdown spectrum of GW150914 suggests the presence of a dominant fundamental $(\ell, |m|, n) = (2,2,0)$ mode and a weaker, short-lived first overtone, the $(2,2,1)$ mode \citep{2019PhRvL.123k1102I,FinchMoore2022,Cotesta+2022,IsiFarrComment2023,Carullo+Reply2023,Ma+2023,Correia+2024}; we include these modes in fits to the post-peak signal using both the generic and aligned-spin ringdown models. We choose the start time of our analysis ($t_0$) to be the peak goecenter time ($t_{\mathrm{peak}}$) corresponding to the maximum likelihood sample of the NRSur7dq4 waveform parameter estimation on GW150914 \citep{NRSur7dq4,NRSurCat1,NRSurCat1_DR}. We use this same sample draw for the right ascension ($\alpha$), declination ($\delta$), and reference polarization angle ($\psi$) to fix the extrinsic parameters of our analysis; the angle $\Delta\psi$ is measured relative to this frame. See Table \ref{tab:inputs} for the input values for these parameters.
\begin{table}[!ht]
  \caption{Extrinsic parameters in the GW150914 analysis.
  % $t_{\mathrm{peak}}$ is the geocenter time corresponding to the peak strain. We begin our analysis at the peak, so $t_0 = t_{\mathrm{peak}}$. The angles $\alpha$, $\delta$, $\psi$ are used to compute the instrument antenna pattern functions $F^{(+/\times)}$ and the geometric time delays between interferometers.
  }
  \label{tab:inputs}
  \begin{ruledtabular}
    \begin{tabular}{cccc}
      $t_0$ (GPS s) & $\alpha$ (rad) & $\delta$ (rad) & $\psi$ (rad) \\
      \hline
      1126259462.4081378 & 2.160164 & $-1.254455$ & 0.675016 \\
    \end{tabular}
  \end{ruledtabular}
\end{table}
We condition the native 16,384 Hz strain data in accordance with the procedures in \cite{2025PhRvD.111d4070S} by targeting an analysis segment with a duration of 0.2 seconds beginning at the time-discretized sample closest to $t_{\mathrm{peak}}$, then downsampling to 2048 Hz using the anti-aliasing digital filter, and high-pass filtering above 10 Hz. 

\subsection{Remnant BH properties} \label{subsec:iiiA}

First, we evaluate the aligned-spin ringdown model's measurements of remnant BH properties. The parameter space spanned by the aligned-spin model forms a subspace of that of the generic model; hence, the two models should, and do, produce the same results for ringdown signals arising from BBH mergers that exhibit equatorial reflection symmetry, as GW150914 appears to. The two ringdown models agree up to sampling noise and potential prior effects; the polarization structure inferred from the IMR analysis is consistent with a region of parameter space in which the models' ellipticity priors disagree (see the Sec. \ref{sec:ii} and App. \ref{subsec:app3} for details), but we show that there is enough polarization information in the ringdown of GW150914 for the generic model inference of remnant properties to reduce to that of the aligned-spin model.
% \mi{this is too much information thrown at the reader without context. a discussion of the difference in priors makes more sense in the previous (background) sections.}
% Where the generic model assigns parameters $\{\epsilon, \theta\}_j$ to each mode $j$, the aligned-spin model parametrizes $\epsilon_j$ by a shared $\iota$ and rotates each mode's polarization ellipse by a common $\Delta\psi$.

\begin{figure}[!ht]
    \centering
    \includegraphics[width=\linewidth]{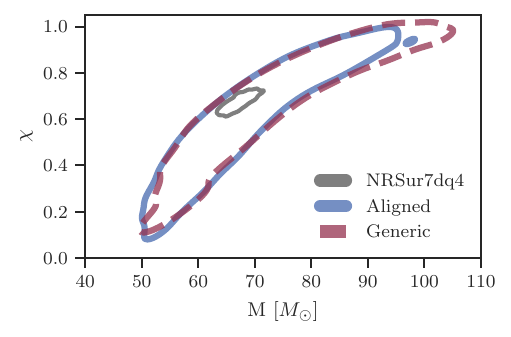}
    \caption{Comparison of $M_f$ and $\chi_f$ posteriors for GW150914 between ringdown and IMR analyses. The results of two ringdown analyses, both fitting the signal with \{(2,2,0),(2,2,1)\}, are shown, with $90\%$ credible contours; the dashed plum contour corresponds to the unconstrained, generic ringdown model, whereas the solid blue contour arises from fitting the signal with the constrained aligned-spin model. The NRSur7dq4 posterior (black) is encapsulated within both ringdown posteriors, suggesting successful recovery of remnant BH parameters and consistency between the ringdown models and the full signal analysis.} 
    \label{fig:GW150914MChi}
\end{figure}

We compare the remnant final mass, $M_f$, and final spin, $\chi_f$, measurements from our aligned-spin ringdown analysis against those of the NRSur7dq4 analysis that fits the entire signal. 
% \mi{do you ever explain how these are produced? i.e., that they are derived from the entire waveform?}.
We consider \textit{consistency} between posteriors, visually assessing the overlap between 90\% credible regions. In Fig. \ref{fig:GW150914MChi}, the aligned-spin ringdown contour (solid blue) entirely encloses the IMR contour (black) at the 90\% credible level, implying consistent $M_f$ and $\chi_f$ measurements between the two analyses. 
Furthermore, the two ringdown polarization models produce consistent posteriors, as expected. The more tightly constrained joint distribution inferred by the aligned-spin model is a reflection of the reduced number of free parameters in the model. Fig. \ref{fig:GW150914MChi} shows that the generic model inference extends to slightly higher masses and spins, a consequence of broader mode amplitude posteriors with greater support for mode-switching: the higher-mass portion of the posterior corresponds to the model assigning the (2,2,0) and (2,2,1) zero and non-zero amplitudes, respectively, such that any polarization structure is measured from the (2,2,1) alone while the (2,2,0) remains unconstrained. The aligned-spin model, on the other hand, enforces that both modes must simultaneously drive a measurement of the signal's ellipticity; even if the (2,2,1) is too quiet to be resolvable above the noise, its ellipticity is not free to vary as in the generic model since it is tied to that of the (2,2,0) mode.
This reduces the potential for mode-switching, thus restricting the remnant and mode parameters to a narrower space.
% Fig. \ref{fig:GW150914MChi} is a good visual representation of the aligned-spin parameters spanning a subspace of the generic model parameters, as the generic model inference (dashed plum) extends to higher masses inaccessible by the aligned-spin model.
% \mi{is there an explanation for this? that's a question you may expect from a referee: why does adding more polarization parameters provide access to higher masses and spins?}

We can statistically assess which of the two polarization models is preferred by the data using the leave-one-out (LOO) cross-validation metric \cite{2015arXiv150704544V}.\footnote{For our ringdown analyses in particular, we prefer to use the LOO for reduced sensitivity to the prior volume---given the limited SNRs of ringdown signals and the lack of physically-motivated priors, we favor broad priors on parameters like $M_f$, $\chi_f$, and $A_{\ell |m| n}$ which might artificially contribute to the Occam penalty in a Bayes factor assessment, for example (see the final paragraph of Sec. 3 in \cite{2019PASA...36...10T}).
% \mi{more briefly: we prefer this because we have no way of setting principled priors on the parameters of the ringdown model and, indeed, we favor them to be broad.}
} 
% \mi{offer a short explanation/intuition for what the LOO is and also maybe point to Harrison's paper for more details} 
LOO estimates the log predictive density for each data point when it is removed from the dataset fit by a given model. Summing over all observations, the LOO quantifies the model's predictive performance via the expected log pointwise predictive density (ELPD); see \cite{2015arXiv150704544V} and Sec. III. C. of \cite{2023PhRvD.108f4008S} for technical details. The \texttt{compare} method in the \textsc{arviz} package \citep{arviz} finds the ELPD difference, $\Delta$ELPD, and its standard error between the best-ranked model (i.e., the model with the strongest predictive accuracy) and the remaining models in the comparison. The ELPD difference measurement is influenced by two main noise sources---the limited number of observations in the dataset and the finite number of samples used to estimate the ELPD difference---so interpreting the strength of the model preference should keep both noise sources in mind.  The effect of the latter noise source is estimated by a standard error on the computed ELPD.  For the former, we can think of the ELPD as a predictive log likelihood, so it is approximately equivalent to $-\frac{1}{2}\chi ^2$ up to an additive constant, and its statistics under repeated sampling with finite observations would be approximately $\chi^2$ with one degree of freedom.  With this in mind, we can translate our predictive accuracy measurements into $|\Delta\chi^2|$ ranges to assess the degree of model preference, and use the estimated standard error to ensure we have sufficient sampling to be confident in these preferences.
% We use the z-score ratio of $\Delta$ELPD to its standard error as a proxy for the strength of the model preference, so a z-score of e.g. ${\sim} 2$ corresponds to a ${\sim} 2\sigma$ preference for the best-ranked model.

A comparison of the generic and aligned-spin ringdown model fits to the post-peak signal of GW150914 demonstrates a preference for the aligned-spin model with $\Delta \mathrm{ELPD} = 1.12 \, \pm \,0.66$, or $|\Delta \chi^2| {\sim} 2.24 \, \pm \,1.32$ suggesting a 1--2$\sigma$ preference. 
% \added[id=NK]{One can think of the ELPD as a predictive log likelihood, so it is approximately equivalent to $-\frac{1}{2}\chi ^2$ up to an additive constant. With the intuition that $\Delta$ELPD is proportional to $\Delta\chi^2$, it is clear that $\Delta\mathrm{ELPD} \gtrsim 4$ is considered a statistically significant, or $\sim 3\sigma$, improvement in predictive accuracy with one model over another \cite{2015arXiv150704544V}.} \mi{this should be moved up; also, why is this clear? do not assume things are clear for the reader unless it's really patently trivial. Where does how does 4 come from $3\sigma$? also, generally we should de-emphasized arbitrary thresholds like this, it's better to give some intuition and say what $\Delta\mathrm{ELPD}$ correspond to $1\sigma$, $2\sigma$, etc.} 
So, the data prefer the aligned-spin model over the generic model but do not offer strong evidence for far better predictive performance with either model. This result suggests that the extra freedoms of the generic model are not necessary to describe the postmerger signal of GW150914, so we learn from the ringdown alone that the system is indeed consistent with obeying equatorial reflection symmetry. This encourages us to leverage the constraints of the aligned-spin model to extract polarization information from the ringdown of GW150914 and infer the inclination of the system, which we do in the following section.

% This result suggests that the aligned-spin model is a slightly better description of the data than the generic model \mi{you say above that the aligned-spin model is a subset of the generic model, so in what sense is it a better description?}. 
We have demonstrated that, from the perspective of both parameter estimation and goodness-of-fit, constraining polarization amplitudes and phases to obey equatorial reflection symmetry serves as an adequate model of ringdown signals from non-precessing systems like GW150914.

\subsection{Polarization analysis} \label{subsec:iiiB}

% Next, we demonstrate our measurement of polarization content using ringdown signal alone.
The symmetry embedded in the aligned-spin model enables a sensitive measurement of the GW polarization structure from the post-peak signal alone, which is tied to a measurement of the system's inclination. Recall from the discussion in Sec. \ref{sec:ii} that we implement a flat prior on $\cos\iota$ in this restricted model.
% \mi{this is fine, but I think some mention of this would make sense above when you introduce the two models. In which case you could reduce this paragraph to a sentence.}
% \mi{generally figures should appear in order; so maybe just make a plot of the priors or point to \cite{Isi2023} since I show that there I think}
\begin{figure}[t]
    \centering
    \includegraphics[width=\linewidth]{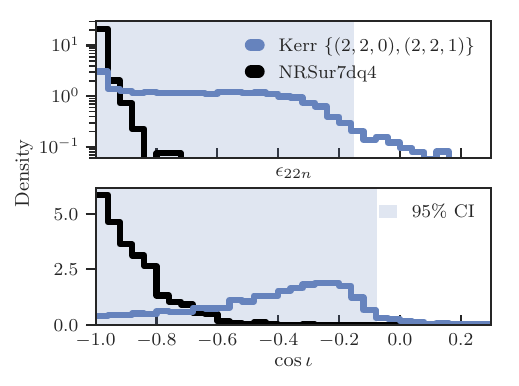}
    \caption{The aligned-spin ringdown model (blue) measures the inclination of GW150914 and thus its polarization content in agreement with the IMR analysis (black). The light blue regions mark the 95\% credible intervals of the post-peak analysis posteriors. \textit{Top:} The ringdown analysis infers the polarization content of GW150914 to be left-handed circular, in support off the IMR result (computed using Eq. \ref{eq:ellip} for $\ell=|m|=2$ modes). \textit{Bottom:} The aligned-spin model measures the inclination of the remnant to be face-off with $97\%$ credibility.}
    \label{fig:GW150914EllipCosi}
\end{figure}
We compare the posteriors for the observed signal ellipticity of GW150914 inferred by the aligned-spin ringdown analysis and the NRSur7dq4 analysis. 
\begin{figure}[b]
    \centering
    \includegraphics[width=\linewidth]{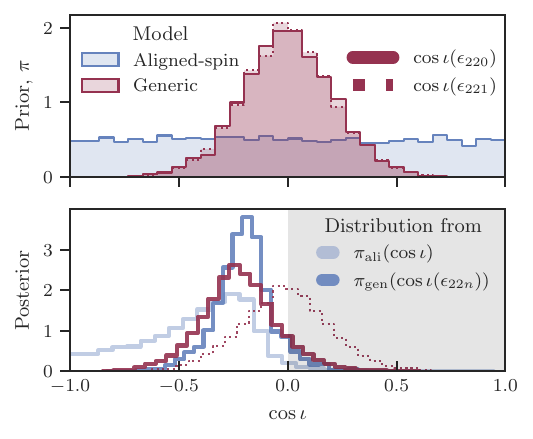}
    \caption{Comparing $\cos\iota$ posteriors between polarization models that fit the ringdown of GW150914 with the Kerr (2,2,0) and (2,2,1) modes. 
    \textit{Top:} The prior on $\cos\iota$ in the aligned (blue) and generic (dark plum) models. There is no explicit $\cos\iota$ parameter in the generic model, so the priors shown are derived using the inverse of Eq.~\eqref{eq:ellip} applied to this one single mode. As discussed at the beginning of this section, the priors disagree at the bounds of the  prior range, with generic model prior having zero support at $\pm 1$. Because both ringdown modes share the same $(\ell, |m|)$, the solid and dotted plum distributions are identical up to sampling noise.
    \textit{Bottom:} 
    The aligned model directly produces a posterior on $\cos\iota$ (light blue, same as the dark blue curve in the bottom panel of Fig. \ref{fig:GW150914EllipCosi}), which is common to all modes in the model. For comparison, we can reweight this posterior to the dark plum generic model effective $\cos\iota$ prior in the top panel to obtain the dark blue curve.
    The solid and dotted curves correspond to the effective $\cos\iota$ computed from the $\epsilon_{220}$ and $\epsilon_{221}$ posteriors, respectively, using Eq.~\eqref{eq:ellip}.
    The shaded gray region highlights the $\cos\iota \geq 0$ portions of the posteriors, which contend with the face-off inference from the IMR analysis. 
    % \mi{the y labels don't match: in both cases you have densities, and the $\pi$ notation is not explained anywhere. it would be more helpful to have "Prior" and "Posterior" written out explicitly.}
    }
    \label{fig:GW150914CosiModelComp}
\end{figure}

IMR parameter estimation produces a posterior on $\cos\iota$, which we can convert to $\epsilon_{22n}$ using Eq.~\eqref{eq:ellip}. In Fig. \ref{fig:GW150914EllipCosi} (top), we show the $\epsilon_{22n}$ posteriors from both analyses,
% which agree on greatest support for $\cos\iota \approx -1$.
which agree on greatest support for left-handed circular polarization, $\epsilon_{22n} \approx -1$.
In making this mode ellipticity measurement with the aligned-spin model, we also infer a physically meaningful quantity, $\cos\iota$, from the ringdown analysis that we cannot necessarily glean directly from the generic model; the bottom panel of Fig. \ref{fig:GW150914EllipCosi} shows a comparison of the posteriors on $\cos\iota$ between the post-peak and IMR analyses. From the ringdown alone, we measure the probability that GW150914 is a face-off ($\cos\iota < 0$) system to be 97\%.

We next demonstrate the sensitivity of the aligned-spin model to polarization information in the ringdown in comparison with the generic model. Because GW150914 is consistent with a non-precessing, face-off system, its
measured QNM parameters \textit{should} lie in the aligned-spin subspace of the
generic polarization ringdown model: each
mode's $\epsilon_{\ell |m| n}$ should be controlled by a common inclination. If this is the case, we expect the mode
ellipticities $\epsilon_{22n}$ measured by the generic model to reduce to those inferred by the aligned-spin model and thus be identical\footnote{Recall that a system obeying equatorial reflection symmetry
produces a ringdown whose mode ellipticities, $\epsilon_{\ell|m|n}$, are
directly parameterized by a common inclination angle through angular functions
that depend only on $\ell$ and $m$.} up to statistical uncertainty.
% , and be consistent between the aligned and generic
% models, which implies consistency between the two modes
% themselves \mi{wut? I think I know what you mean but this center is hard to parse, you are using ``consistency'' in two different senses---help the reader! we are not in your head :) };
% \footnote{Recall that a system obeying equatorial reflection symmetry
% produces a ringdown whose mode ellipticities, $\epsilon_{\ell|m|n}$, are
% directly parameterized by a common inclination angle through angular functions
% that depend only on $\ell$ and $m$.}
We can use the inverse of Eq.~\eqref{eq:ellip} to numerically compute induced inclination angles from
the generic model mode ellipticies, assuming that the system
exhibits equatorial reflection symmetry. 
% \mi{do you also reweight the priors?}

In the top panel of Fig.~\ref{fig:GW150914CosiModelComp}, we show the explicit and implicit priors on $\cos\iota$ in the two polarization models: $\pi_{\mathrm{ali}}$ for the aligned model (blue) and $\pi_{\mathrm{gen}}$ for the generic model (plum).
We compare the actual and induced $\cos\iota$ posteriors from the aligned and generic model fits, respectively, in the bottom panel of Fig.~\ref{fig:GW150914CosiModelComp}. The $\cos\iota(\epsilon_{22n})$ posteriors from the generic polarization analysis overlap but peak at different values; while we expect the effective $\cos\iota$ to be identical for (2,2,0) and (2,2,1) in the limit of infinite detector sensitivity, noise interacts with the different time-dependence of the ringdown modes to add uncertainty onto their measurements in line with what we see in Fig.~\ref{fig:GW150914CosiModelComp}. Furthermore, the overtone posterior is nearly identical to its prior, implying the SNR is too low to constrain polarization structure from the sub-dominant mode. It is therefore reasonable that modest SNR, as well as imperfect spin-orbit alignment, account for slight discrepancies in measurements of $\epsilon_{220}$ and $\epsilon_{221}$ by the generic model.

Since the generic model infers a better constrained ellipticity for the $(2,2,0)$ mode, we use it to compute the induced $\cos\iota(\epsilon_{220})$ and compare with the $\cos\iota$ posterior from the aligned-spin model in the bottom panel of Fig.~\ref{fig:GW150914CosiModelComp}. For a robust comparison, we reweight the aligned-spin posterior (light blue) to the generic model's effective $\cos\iota$ prior, $\pi_{\mathrm{gen}}$, to obtain the dark blue distribution. 
% There is almost no difference between the blue curves, before and after reweighting, suggesting that the measurement is likelihood-dominated with non-zero support at $\epsilon=-1$ despite lacking prior density in that region. 
This reweighted $\cos\iota$ measurement (dark blue) and the ``effective'' $\cos\iota(\epsilon_{220})$ (dark plum) posterior are consistent with one another; however, only 89\% of the latter's probability mass lies below zero and supports the face-off inference from the IMR analysis, as opposed to 93\% of the former's. In summary, the aligned-spin model recovers GW150914's polarization structure, and in doing so it constrains the inclination of the system to be face-off; the generic model makes a marginally less precise measurement of the polarization content that agrees with the IMR analysis using just the dominant mode and cannot simultaneously translate the polarization of additional modes into an overall $\cos\iota$ measurement. 

% considering it does so in agreement with the IMR analysis from only the $(2,2,0)$ mode, and is thus unable to translate all  polarization to the system's orientation at all.
% \mi{except that's not true because you just did that! you just turned a measurement of $\epsilon_{220}$ into a measurement of $\cos\iota$. need to rephrase.}
\begin{figure*}[ht]
    \includegraphics[width=\textwidth]{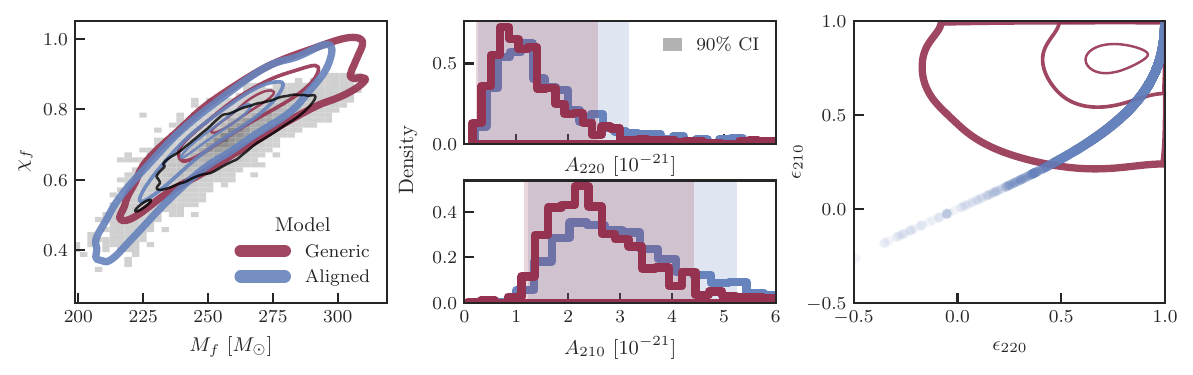}
    \centering
    \caption{Various parameter posteriors from ringdown fits with both polarization models to GW190521 data. \textit{Left:} Remnant BH $M_f$--$\chi_f$ posteriors, with the (90, 50, 10)\% credible contours ordered from thick to thin depicted for the two models. The gray histogram represents the full NRSur7dq4 posterior, with the 90\% credible contour overlayed in black. Both polarization models measure the remnant properties in agreement with the IMR analysis at the ${\sim}90\%$ credible level. \textit{Center:} Posteriors on the (2,2,0) and (2,1,0) mode amplitudes, with shaded regions marking the $90\%$ credible intervals. \textit{Right:} Mode ellipticity posteriors, reweighted to a uniform prior for the generic model. The contours mark the (90, 50, 10)$\%$ credible regions from thickest to thinnest. The aligned-spin model posteriors form a curve parameterized by the inferred $\iota$.}
    \label{fig:GW190521real_comp}
\end{figure*}

\section{Precessing test case: GW190521} \label{sec:iv}

% \begin{figure}[t]
%     \centering
%     \includegraphics[width=\linewidth]{figs/prec/GW190521_220210_inj_samp.pdf}
%     \caption{We choose a sample draw from the posterior of the fit shown in Figure \ref{fig:GW190521_fit_mchi}, the corresponding $M$ and $\chi$ samples indicated by the dashed lines. We use this sample to build a ``GW190521-l ike'' damped sinusoid injection.} 
%     \label{fig:GW190521_fit_mchi_sample}
% \end{figure}

Precessing systems, i.e., BBHs with one or both component spins misaligned with the binary's orbital angular momentum, do \textit{not} obey the equatorial plane symmetry imposed within the aligned-spin ringdown model. Inaccurate assumptions about the polarization structure of GWs from such sources can bias the measured ringdown model parameters. To investigate the extent of this bias, we compare arbitrary- and restricted-polarization ringdown analyses of, first, a real GW event with evidence of precession, followed by a set of
synthetic signals designed to mimic the polarization characteristics of precessing systems. 
% We expect the aligned-spin and generic model inference to disagree for ringdown signals from truly precessing systems that breaks equatorial reflection symmetry.

% We construct these injections from a representative sample drawn from the posterior of a generic model fit to the ringdown of GW190521 \citep{GW190521}.
We choose to focus on GW190521 \citep{GW190521} for our real data analyses because of evidence in the literature that its source is consistent with exhibiting precession from spin-orbit misaligment \citep{GW190521,GW190521props,Olsen+2021,Biscoveanu+2021,2023PhRvD.108f4008S,Miller+2024}. Spin misalignment at merger imprints itself on the ringdown through greater excitation of $\ell\neq |m|$ multipoles like the (2,1,0) mode \citep{Berti+2007,2023PhRvD.108f4008S,Zhu+2025}; additionally, the polarization structure encoded by these modes will generally be inconsistent with the restrictions of the aligned-spin ringdown model.
% While other astrophysical interpretations of the source of GW190521 have been considered, these imply ringdown mode content that does not necessarily break the equatorial reflection symmetry. So, to ensure that our synthetic signals are representative of the polarization structure of a signal from a precessing system and that our ringdown inference is not degenerate with the choice of modes we include in our model, we build the injections from a reconstruction of a ringdown-only analysis of real GW190521 data that supports the precession hypothesis. \mi{this logic is convoluted, going round and round\dots just first say that GW190521 can be fit with a 220+210 combination that does not satisfy the aligned symmetry and that we take the fit from Siegel+ as a reference for an example of a precessing ringdown.} 
Indeed, we fit the ringdown of GW190521 with a template including the Kerr (2,2,0) and (2,1,0) with and without restriction on their polarization amplitudes and phases.
In particular, we reproduce the Siegel et al. \citep{2023PhRvD.108f4008S} analysis which shows that fitting the ringdown of GW190521 with the generic model including two damped sinusoids labeled as the Kerr (2,2,0) and (2,1,0) modes produces remnant mass and spin posteriors that are consistent with those inferred by NRSur7dq4, which is calibrated to NR simulations that include the effects of spin-orbit misalignment \citep{NRSur7dq4}. This analysis demonstrates self-consistent and IMR-consistent remnant BH mass and spin measurements for fits starting as early as 6.35 milliseconds (${\sim}5 t_{M_f}$, where $t_{M_f} = G(1+z)M_f/c^3$) before the reference peak of the signal, or roughly $2\sigma$ before the median peak strain time measured by NRSur7dq4.

Then, to more carefully explore how the limitations of the aligned-spin model for ringdown signals from precessing systems manifest in ringdown inference, we compare the two polarization models' performance when fitting GW190521-like synthetic signals of varying SNR. Taking the reconstructions from the ringdown fits to the real GW190521 data as references, we create the simulated injections from representative posterior draws. We demonstrate that inconsistencies in the inferred distributions of remnant BH and mode properties between the two models may provide evidence of precession from the ringdown alone.

\subsection{Fitting real GW190521 data} \label{subsec:iv_realdata}
\begin{figure}[ht]
    \includegraphics[width=\linewidth]{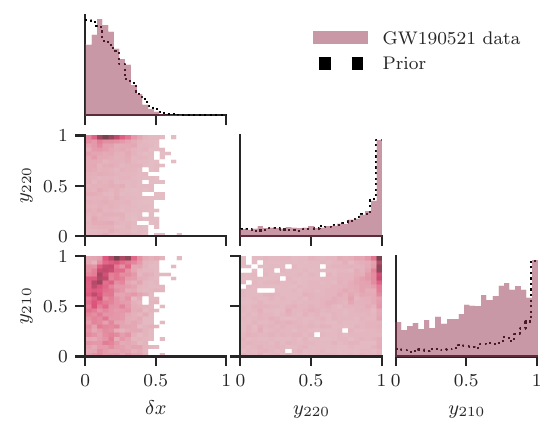}
    \centering
    \caption{Posteriors (priors) in solid pink (dotted black) on parameters that quantify deviations from the aligned-spin parameter subspace. Support at zero corresponds to a posterior consistent with the polarization structure restrictions of the aligned-spin model. The $\delta x$ distribution describes how strongly the inclination angle of the system can be tied to the measured mode ellipticities, with $\delta x = 0$ representing $\epsilon_{\ell |m| n}$ that is derived directly from $\iota$ as in Eq.~\eqref{eq:ellip}. The $y_{\ell |m| n}$ distributions describe how well the intrinsic mode amplitudes obey the equatorial reflection symmetry constraint, $C_{\ell -|m| n} = (-1)^\ell C^*_{\ell |m| n}$. The $\delta x$ posterior peaks slightly away from zero. While the $y_{210}$ and $y_{220}$ distributions are poorly constrained, the $y_{210}$ posterior is driven away from the prior, suggesting a more likelihood-dominated measurement. See Fig. \ref{fig:B4_alisubspace} for a comparison of these posteriors when the underlying polarization structure of the signal is known a priori to obey or break equatorial reflection symmetry.}
    \label{fig:GW190521_alisubspace}
\end{figure}

We first analyze the ringdown of GW190521 using the Kerr (2,2,0) and (2,1,0) mode combination, following \cite{2023PhRvD.108f4008S}, with both the generic and aligned-spin polarization models. We apply the same conditioning to the data as described in Sec.~\ref{sec:iii} for our GW150914 analysis and target an analysis start time of $t_0 = t_{\mathrm{peak}} - 5t_{M_f}$,  
% \mi{so you do this before the peak?? why?? this is inviting pushback} 
where $t_{M_f} = 1.23 \mathrm{~ms}$, and $t_{\mathrm{peak}}$ and $M_f$ are determined by the sample draw corresponding to the median Livingston peak time inferred from the full signal using NRSur7dq4 \citep{GWTC3}. We start the analysis before the peak estimate, motivated by the early-time fits to the ringdown-dominated signal in \cite{2023PhRvD.108f4008S} and its broad peak time distribution \citep{GWTC3}, to leverage the most SNR. The fixed analysis input parameters are listed in Table \ref{tab:inputs_GW190521}.
\begin{table}[!bt]
  \caption{Extrinsic parameters in the GW190521 ringdown analyses.
  % $t_0$ is the geocenter time corresponding to the target start time of the $0.2\,\mathrm{s}$ analysis segment in each detector. We begin the fit slightly before the peak: $t_0 = t_{\mathrm{peak}} - 5 t_{M_f}$.
  }
  \label{tab:inputs_GW190521}
  \begin{ruledtabular}
    \begin{tabular}{cccc}
      $t_0$ (GPS s) & $\alpha$ (rad) & $\delta$ (rad) & $\psi$ (rad) \\
      \hline
      1242442967.4027324 & 0.043307 & $-1.162271$ & 2.218851 \\
    \end{tabular}
  \end{ruledtabular}
\end{table}

We show the results of the GW190521 ringdown  analyses with the generic and aligned-spin models in Fig.~\ref{fig:GW190521real_comp}. The posteriors on remnant mass and spin, mode amplitudes, and mode ellipticities inferred by the two polarization models are largely consistent with one another at the $90\%$ credible level. Although there is a slight shift in the joint $M_f$--$\chi_f$ distribution measured by the aligned-spin model with respect to that of the generic model, the bulks of the posteriors agree. Both models also infer nearly identical posteriors on the amplitude of each mode, with both modes required to be present (non-zero amplitudes). To assess the polarization structure measured by both models, we reweight the mode ellipticity posteriors of the generic model to a uniform prior (see the discussion of prior differences between the models in Sec.~\ref{sec:ii}, as well as more motivation for this reweighting in Sec. \ref{subsec:ivA} and App. \ref{subsec:app3}). The aligned-spin model more tightly constrains the inferred polarization structure of the signal, as the bulk of the ellipticity posteriors is encompassed by those of the generic model at the $90\%$ credible level. 

Recalling the aligned-spin subspace deviation parameters introduced in Sec.~\ref{sec:ii}, $\delta x_k$ and $y_{\ell |m| n}$, we can further assess the consistency of the two polarization models' ringdown inference by the degree to which the generic fit posterior supports the symmetry of the aligned-spin model. In Fig.~\ref{fig:GW190521_alisubspace}, we show the posteriors on the aligned-spin subspace deviation parameters for the generic model fit, as well as the priors given by the dotted black marginals. While the inferred distributions for all three parameters have support at zero, representing the aligned-spin model constraints and consistent with their respective priors, they also have some support for departures from the polarization restrictions imposed by equatorial reflection symmetry. The $\delta x$ posterior exhibits a peak slightly away from zero. Furthermore, while neither $y$ parameter is especially tightly constrained given the modest SNR of GW190521 and the $y_{220}$ posterior is functionally identical to its prior, the inferred $y_{210}$ distribution is distinct from the shape of its prior. This suggests there is enough polarization information in the post-merger signal of GW190521, particularly in the 210 mode, to hint at structure that violates the intrinsic mode amplitude relationship enforced by the aligned-spin model, but not enough to discount the validity of the aligned-spin constraints. Later in Sec.~\ref{subsec:ivC}, we compare these posteriors when the source of the underlying signal is known a priori to obey or to break equatorial reflection symmetry. 

The inferred aligned-spin deviation parameters suggest that there is polarization content in the ringdown of GW190521 that is accessible by the generic model due to its polarization freedoms but not by the aligned-spin model. Despite this, due to the limited available SNR, the comparison of the polarization models' inference results shown in Fig.~\ref{fig:GW190521real_comp} does not demonstrate a strong bias indicative of the aligned-spin model poorly describing the data. Furthermore, comparing the two models using the LOO, we find a 1--2$\sigma$ preference for the aligned-spin model with $\Delta\mathrm{ELPD} = 1.66 \pm 0.8$, or $|\Delta\chi^2| \sim 3.32 \pm 1.6$. This is a modest preference at best, suggesting there is not enough evidence from the data to rule out the aligned-spin model as a poor description.

For the modest SNR of GW190521 and the particular noise realization in its data, it is not clear from its ringdown polarization inference alone that GW190521 breaks equatorial reflection symmetry. This is consistent with the conclusions of the time-domain dissection study detailed in \cite{Miller+2024}, which show that the isolated merger-ringdown portion of GW190521 data does not demand precession. To more robustly explore how ringdowns from precessing systems may suffer biased inference by incorrectly modeling their polarization amplitudes and phases, we use both polarization models to analyze a set of injections that encode GW190521-like polarization content. We construct these injections with SNRs comparable to or greater than that of GW190521 in order to forecast the consequences of polarization mismodeling in future observing runs with improved detector sensitivity. 

\subsection{Constructing ``GW190521-like'' injections} \label{subsec:ivA}

% \mi{so after all that we actually start with the actual 190521, not the injections? this needs to be restructured: (1) GW190521 is well fit by 220+210, are those recovered amplitudes consistent with the aligned subspace? No or poorly; then (2), what happens if you fit 190521 with the aligned-spin model? it should be biased or fit not as well; (3) simulated injections}

\begin{figure}[b]
    \centering
    \includegraphics[width=\linewidth]{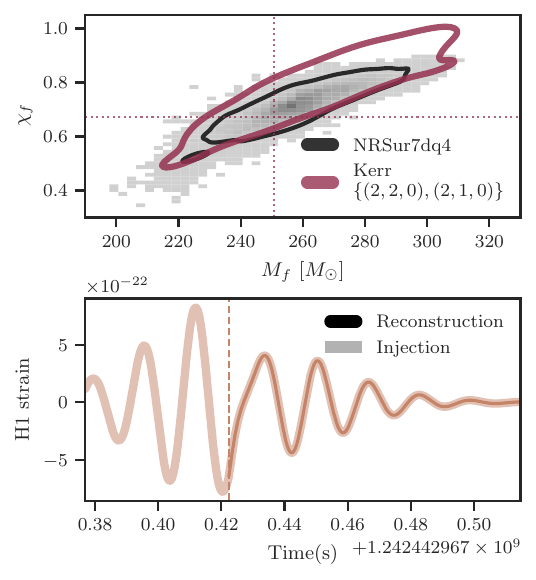}
    \caption{
        \textit{Top:} Generic ringdown model fit to the real GW190521 data, using Kerr modes \{(2,2,0), (2,1,0)\}. Both $90\%$ credible contours and the histogram are equivalent to those in the leftmost panel of Fig. \ref{fig:GW190521real_comp}. The resulting inference (plum) is consistent with the IMR $M_f$---$\chi_f$ measurements (black), so the fit's waveform reconstructions should retain the elliptical polarization structure of the signal. We draw a sample from the post-peak analysis posterior, indicated by the dotted plum lines, to build a ``GW190521-like'' damped sinusoid injection.
        \textit{Bottom:} Detector-frame damped sinusoid injection parameterized by the chosen posterior sample draw from the top panel (we project the signal into all three LIGO and Virgo interferometers, showing just the Hanford projection here). We confirm that the injected waveforms (thicker, lighter) match the detector reconstructions obtained from the draw in the top panel. The vertical dashed line indicates the start of the analysis segment for both the real data and injection analyses.} 
    \label{fig:GW190521_fit_mchi}
\end{figure}

In the previous section, we fit the ringdown portion of the real GW190521 data with a QNM model that allows posterior support for equatorial asymmetry; we draw a sample from this posterior to parameterize synthetic signals representative of the polarization structure of GWs from a precessing system.
% Assuming the Kerr metric, we follow the analysis in \cite{2023PhRvD.108f4008S} and fit the postmerger GW190521 data using the Kerr $\{(2,2,0), (2,1,0)\}$ modes and generic polarizations with no additional constraints on QNM amplitudes and phases.
% From$\,_sY_{\ell m}(\theta, \varphi)$ symmetry selection rules \citep{Berti+2007}, we know that odd-parity multipoles, and subsequently QNMs, are suppressed in the harmonic decompositions of GWs from non-precessing mergers. The $(2,1,0)$ mode may instead be appreciably excited in remnants that exhibit spin-orbit misalignment \citep{Zhu+2025}, and including it in our model imposes posterior support for equatorial asymmetry. There is also evidence in the literature of a dominant $(2,1,0)$ mode in the ringdown spectrum of GW190521 \cite{2023PhRvD.108f4008S} and perhaps GW231123 \mi{cite detection paper and harrison's paper}. 
The result of the generic model Kerr $\{(2,2,0),(2,1,0)\}$ fit to GW190521 is shown in the top panel of Fig.~\ref{fig:GW190521_fit_mchi} in plum, demonstrating consistency with the IMR analysis through the near-total overlap of the two sets of 90\% credible regions. 
% \mi{this paragraph is structured backwards IMO and therefore hard to follow; also there's no motivation for what you are doing: explain first that you are trying to construct a synthetic signal (you didn't do this for 150914)}

\begin{figure*}
    \centering
    \includegraphics[width=\linewidth]{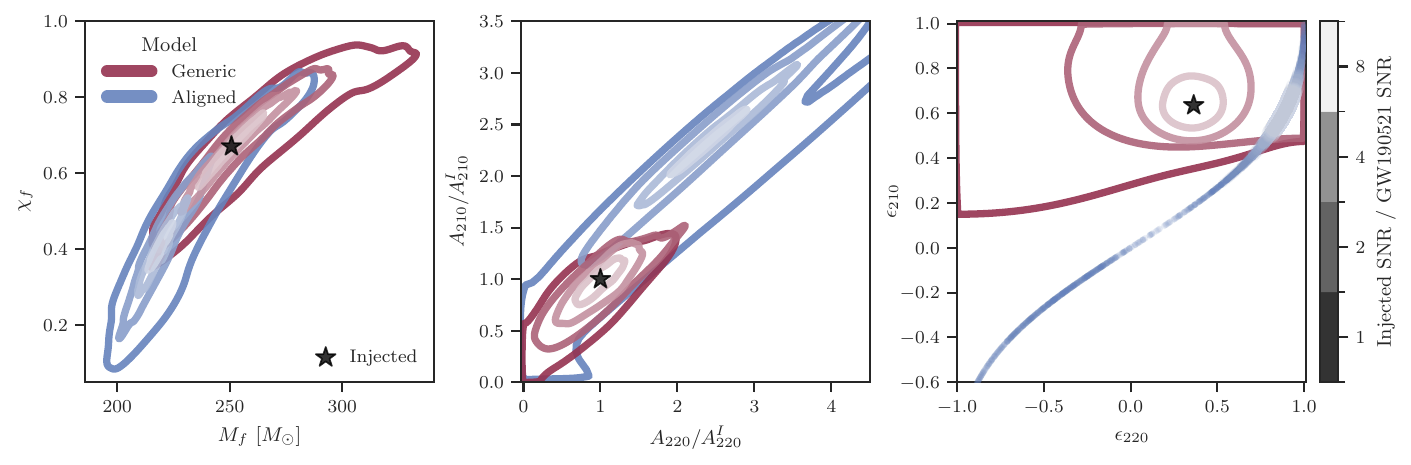}
    \caption{
        Various parameter posteriors from ringdown fits with both polarization models to GW190521-like reflection-asymmetric injections at varying SNRs. All contours mark 90\% credible regions, and the different shades correspond to different values of $B$, the SNR multiplier for a given injection. In all panels, the generic model behaves as expected, more evidently so as $B$ increases: its posteriors center on the truth. 
        \textit{Left:} Remnant BH $M_f$---$\chi_f$ posteriors. The aligned-spin model confidently recovers the injected values for $B=1$, then produces biased measurements for larger $B$.
        \textit{Center:} Posteriors on the ratio of inferred to injected ($A_{\ell|m|n}^I$) mode amplitudes. The aligned-spin model overestimates both mode amplitudes.
        \textit{Right:} Mode ellipticity posteriors, reweighted to a uniform prior for the generic model and otherwise unaltered for the aligned-spin model. The latter posteriors form a parametric curve because they are deterministic functions of the inclination angle. The points along this curve follow the same color conventions as the contours, with lighter shades of blue (and larger markers) corresponding to greater injection SNR. The aligned-spin model produces bimodal measurements for $B=1$ and overestimates $\epsilon_{220}$ for $B \geq 2$.
        % \mi{it would be better to label the colorbar by SNR instead of $B$; to a reader that is skimming the paper $B$ doesn't mean anything. If you want to retain the multiplier, I would label it explicitly as ``Injected SNR / GW190521 SNR''.}
        Compare with the posteriors in Fig.~\ref{fig:injAli_comp} from fits to similar reflection-\textit{symmetric} injections.
        }
    \label{fig:injSNR_param_model_comp}
\end{figure*}

We draw a sample from the full posterior that is representative of the polarization structure of the signal (see App.~\ref{subsec:app3} for more detail), shown in the top panel of Fig.~\ref{fig:GW190521_fit_mchi} in the slice of remnant mass and spin. We build an injection consisting of a damped sinusoidal ringup and a ringdown that replicates the sample draw waveform reconstruction, such that we can still condition the entire signal according to \cite{2023PhRvD.108f4008S} and then discard the data at times prior to the peak in our analysis to minimize any conditioning contamination of the post-peak signal \cite{2021arXiv210705609I}. The complex strain of this synthetic signal $s_{\mathrm{ifo}}(t) = \sum_{n} \big[ F^+ h_{n, \mathrm{inj}}^+ + F^{\times} h_{n, \mathrm{inj}}^{\times} \big]$, before projecting into the detector frame using the same antenna patterns as those of the real GW190521 analysis, takes the form
\begin{align}
    \label{eq:inj}
    h_{n, \mathrm{inj}}(t) &= \frac{1}{2} A_n e^{-\gamma_n|t - t_0|} \quad \times \\ 
    &\Big[(1 + \epsilon_n) e^{-i(\omega_n|t - t_0| - \phi_n + \theta_n)} \nonumber \\
    &+ (1 - \epsilon_n) e^{i(\omega_n|t - t_0| - \phi_n - \theta_n)} \Big] \nonumber \,  ,
\end{align}
following Eq. (8) in \cite{2021arXiv210705609I} with $n=2$ (a damped sinusoid corresponding to the parameters of each of the $(2,2,0)$ and $(2,1,0)$ modes). The value of each mode parameter indexed by $n$ is determined by the posterior draw shown in the top panel of Fig.~\ref{fig:GW190521_fit_mchi}.
% We fit a set of noisy injections at varying SNRs with both the aligned-spin and generic ringdown models. We construct these injections from a posterior sample draw of a generic model fit to the ringdown of GW190521 (the same Kerr \{220+210\} analysis as in \textbf{Harrison 190521}). 

We construct a 4 s time series centered on $t_0$, sampled at 2048 Hz. We inject the signal from Eq.~\eqref{eq:inj}, scaled by a factor to vary the SNR, into the discretized time array and project it onto each interferometer given fixed extrinsic parameters. The resulting data segments per detector are $d_{\mathrm{ifo}}(t) = Bs_{\mathrm{ifo}}(t)$, where $s_{\mathrm{ifo}}(t)$ is the projected injection and $B \in \{1, 2, 4, 8\}$ is the SNR multiplier. The bottom panel of Fig.~\ref{fig:GW190521_fit_mchi} shows the post-peak Hanford reconstruction from the top panel fit plotted over the $B=1$ injection; the reconstruction and injection match, and the peak of $d_{\mathrm{H1}}(t)$ occurs at the same $t_0$ as the target analysis start times of the $\{(2,2,0), (2,1,0)\}$ fit in the top panel (the same is also true for the Livingston and Virgo reconstructions and injections). For each interferometer, we model the instrument noise using an autocorrelation function (ACF) obtained by inverse Fourier transforming the power spectral density (PSD) estimated from the GW190521 data using Welch's method, since the ringdown signal is so short-lived and modestly loud. We use these ACFs for the analysis of GW190521's ringdown in the top panel of Fig.~\ref{fig:GW190521_fit_mchi}, as well as for the analysis of the GW190521-like $s(t)$ that we inject into zero noise.

\subsection{Reflection-asymmetric injections} \label{subsec:ivB}

We assess the aligned-spin ringdown fits to our reflection-asymmetric zero-noise damped sinusoid injections at various SNRs from two perspectives: $1)$ accuracy of injected parameter recovery, and $2)$ data-driven model preference via the LOO. For parameter recovery, we consider remnant mass and spin, QNM amplitudes, and the signal ellipticity itself. 

% \begin{figure}[b]
%     \centering
%     \includegraphics[width=\linewidth]{figs/prec/GW190521_injAll_220210_mchi_inj2_FINAL.pdf}
%     \caption{} 
%     \label{fig:injSNR_mchi}
% \end{figure}
% \begin{figure*}
%     \centering
%     \includegraphics[width=\linewidth]{figs/prec/GW190521_injAll_220210_A_comp_inj2_FINAL.pdf}
%     \caption{}
%     \label{fig:injSNR_A_model_comp}
% \end{figure*}
First, we compare the performance of the aligned-spin and generic models in $M_f$--$\chi_f$ space in the left panel of Fig.~\ref{fig:injSNR_param_model_comp}. As we would expect, the generic model posterior bulk centers on the injected truth, with the $90\%$ credible contours shrinking by a factor of about 2 in each dimension as $B$ doubles for each injection. On the other hand, the aligned-spin model supports the truth at the original GW190521-like SNR ($B=1$), but it produces strongly biased results to lower masses and spins for louder signals.
%\added[id=MI]{This demonstrates that assuming a polarization model that is too restrictive can lead to biased measurements of $M_f$--$\chi_f$.}
At all SNRs, there is a systematic shift between the mass and spin measurements of the two polarization models. At present detector sensitivity, incorrect assumptions about the signal's polarization content and the system's spin-orbit alignment might not affect the remnant BH mass and spin measurement. However, future louder detections will demand closer attention to polarization parameterization. In fact, even with current detector sensitivity, the loudest detections may already be affected by polarization mismodeling.
% \mi{this is one of your major findings, do not underplay it!}

%%% WHAT IS THE SNR OF THE ORIGINAL INJECTION??? %%%

% One of the main goals of BH spectroscopy via ringdown analysis is to measure the remnant BH properties, particularly $M_f$ and $\chi_f$. Typically, we discount QNM models that produce fits with $M_f$---$\chi_f$ posteriors that disagree with the final mass and spin measured by the IMR analysis. Checking the aligned-spin ringdown model's recovery of $M_f$ and $\chi_f$ is thus an important starting point in assessing the model's performance. In Figure \ref{fig:}

% \begin{figure*}
%     \centering
%     \includegraphics[width=\linewidth]{figs/prec/GW190521_injAll_220210_ellip_comp_likelihoods_inj2_FINAL.pdf}
%     \caption{}
%     \label{fig:injSNR_ellip_comp_l}
% \end{figure*}

A physically inaccurate ringdown model can still produce seemingly unbiased measurements of $M_f$--$\chi_f$ if it infers other parameters like mode amplitudes and relative phases with enough flexibility. For this reason, it is important to check the amplitudes recovered by our models. In the center panel of Fig.~\ref{fig:injSNR_param_model_comp}, we show joint posteriors on the ratios of the inferred to injected mode amplitudes, so that the truth lies at $(1,1)$. The generic model expectedly recovers the true amplitudes for all $B$; the aligned-spin model fails to recover the injected values to within the $90\%$ credible interval for any $B > 1$, systematically overestimating both mode amplitudes. This result is consistent with the threshold at which the aligned-spin model breaks in accuracy of $M_f$--$\chi_f$ measurements: the systematic bias exceeds the statistical uncertainty at this credible level in both $M_f$--$\chi_f$ and amplitude space simultaneously. Furthermore, with increasing $B$, the two models' posteriors become increasingly inconsistent; for $B=1$ the bias arises in the bulk of the distributions despite the two models' posteriors agreeing at the $90\%$ credible level, by $B=4$ the $90\%$ credible regions of the two models no longer overlap, and by $B=8$ the posteriors of the two models are entirely inconsistent with one another.
% \mi{there should be some mention of the fact that the bias is still there for $B=1$, it's just that it's below the 90\% CL---but it can be clearly seen in the bulk of the distribution.}
% Analyzing a real signal, with no a priori knowledge of ``truth,'' this inconsistency at high SNRs would nonetheless lead us to believe that the signal contains polarization content that is inaccessible by the aligned-spin ringdown model, i.e., mode excitations that are asymmetric about the equatorial plane of the remnant.

For real data, for which we lack knowledge of the true polarization content of the underlying signal, the inconsistency between the two models' inference of remnant BH properties and ringdown mode amplitudes would be a smoking gun for evidence of precession. In fact, the measurement of a damped sinusoid consistent with the Kerr (2,1,0) mode with the generic model is not sufficient evidence, as it may also be appreciably excited relative to the (2,2,0) in non-precessing systems with unequal mass ratios \citep{Zhu+2025}. 
% So we can leverage the constraints of the aligned-spin model to learn about the dynamics of the progenitor BBH encoded in the polarization structure of the ringdown signal, despite the source of the signal breaking the symmetry of the ringdown model.
So despite the aligned-spin model serving as a poor description of signals from precessing systems, we can leverage its constraints to learn about the dynamics of progenitor BBHs just from the polarization structure encoded in their remnants' ringdowns.
% \mi{this is not very clearly stated: are you trying to make the point that we can use the inconsistency between the two models to alert us to the fact that the signal was precessing? if so, this should be stated more forcefully since this is another key point: we can infer precession directly from the polarization structure of the ringdown (besides the fact that we saw the 210 in the first place, which might also come from asymmetric mass ratios---in fact, at some point it'd be good to check whether the two scenarios differ in terms of their polarization structure---but not necessarily in this paper).}

\begin{figure}[b]
    \centering
    \includegraphics[width=\linewidth]{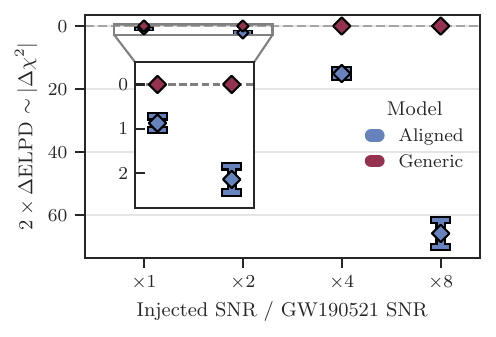}
    \caption{For the set of GW190521-like reflection-asymmetric injections, we compare the generic and aligned-spin models' predictive performance using the LOO. We plot on the ordinate the ELPD difference of both polarization models from the highest-ranked model (best predictive accuracy), scaled by a factor of 2 to draw an analogue to $\Delta\chi^2$ (see Sec.~\ref{subsec:iiiA} for more details). The highest-ranked model lies along zero on the ordinate, and the error bars are the standard error on the measured difference in predictive accuracy. For all $B$, there is at least a moderate preference for the generic model over the aligned-spin model.}
    \label{fig:prec_LOO_comp}
\end{figure}

\begin{figure*}
    \centering
    \includegraphics[width=\linewidth]{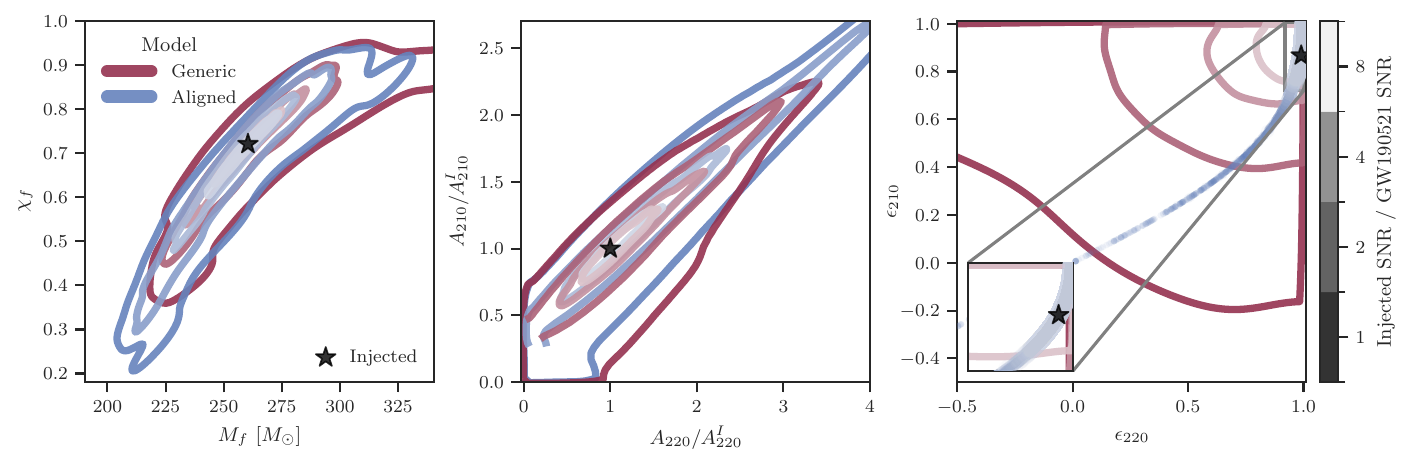}
    \caption{See Fig.~\ref{fig:injSNR_param_model_comp} caption for figure conventions. Various parameter posteriors from ringdown fits with both polarization models to GW190521-like
    % \textit{circularly} polarized
    reflection-symmetric injections at varying SNRs.
    \textit{Left:} Posteriors on remnant mass and spin. The two models confidently recover the injected values for all SNRs. The two models' posteriors also agree at each SNR. We expect these results as the aligned-spin model is a valid description of the injected signals, and the generic model contains the aligned-spin model within its free polarization parameterization.
    \textit{Center:} Posteriors on the ratios of inferred mode amplitudes to the injected values. Both models recover the truth while also maintaining consistency with one another.
    \textit{Right:} Posteriors on mode ellipticities. For all $B$, both models measure $\epsilon$ in agreement with the injected values. The distributions inferred by the aligned-spin model are more tightly constrained than those of the generic model.}
    \label{fig:injAli_comp}
\end{figure*}

Finally, we assess the ringdown models' recovery of the polarization structure itself via measurements of the signal's ellipticity. Recall that the restrictions imposed in the aligned-spin model require that the mode ellipticities $\epsilon_{\ell|m|n}$ are derived from a shared parameter, $\iota$. In the right panel of Fig.~\ref{fig:injSNR_param_model_comp}, we compare the inferred distributions for $\epsilon_{\ell|m| n}$ by both polarization models, reweighted to uniform priors in this quantity for the generic model. We motivate this reweighting in App. \ref{subsec:app3}; its purpose is to restore support at the bounds $\epsilon = \pm 1$ where the generic model priors have zero density. Again, the generic model succeeds in recovering the injected polarizations, while the aligned-spin ringdown model fails to do so for $B\geq 2$. In Sec. \ref{subsec:ivC}, we show that both polarization models would indeed recover the true signal and remnant parameters if the source of the underlying signal was indeed symmetric under equatorial reflection. Interestingly, the aligned-spin model produces biased measurements of the $(2,2,0)$ ellipticity for the reflection-asymmetric injections, but it correctly infers $\epsilon_{210}$ for all $B$. From Eq.~\eqref{eq:ali_pols}, we know the $|m|=1$ mode template has zero amplitude for face-on/off ($\iota=0,\pi$) systems; but the injected $(2,1,0)$ is more than 1.5 times louder than the injected $(2,2,0)$, likely driving the $\cos\iota$ measurement.

In ringdown data analysis for real events, we do not have access to the underlying true properties of a signal;
% ; it is useful to determine some sort of quantitative metric for determining when the aligned-spin model is valid. Considering the posteriors in Fig.~\ref{fig:injSNR_param_model_comp} without knowledge of the injected parameter values, we might naively infer that the aligned-spin ringdown model is an adequate description of the signal when it is as loud as twice the SNR of GW190521. 
it is therefore helpful to have a data-driven metric that indicates the validity of the polarization assumptions made in the aligned-spin ringdown model. To this end, we once again utilize LOO cross-validation with the \texttt{compare} method in \textsc{arviz} \citep{arviz} to compute $\Delta$ELPD between the two polarization models in Fig.~\ref{fig:prec_LOO_comp}. The generic model (dark plum) is the best-ranked model, or the model with the strongest predictive accuracy. This preference is moderate for the lowest SNRs (1--2$\sigma$ by the $\Delta\chi^2$ proxy), and for $B\geq 4$ the generic model is confidently preferred over the aligned-spin model (blue). The goodness-of-fit assessment implies that the restrictions of the aligned-spin model pose an insufficient description of the polarization structure in these signals, indicating support for a precessing source even at currently achievable SNRs. The significance of this model comparison complements the inconsistent posteriors inferred by the two models as discussed previously in this section. 

\subsection{Validation with reflection-symmetric injections} \label{subsec:ivC}
% \mi{doesn't this belong in the previous section where we studied non-precessing signals? I guess you choose to place it here simply because you used GW190521-like parameters. But the structure is currently a bit skewed: we analyze actual 150914 data for the aligned case, but then for the non-aligned case we start with synthetic signals, never analyze actual 190521 data and then do synthetic aligned injections. The easiest fix might be to add some discussion of the actual 190521 data and then maybe forward reference this subsection above from the aligned section}

Here we repeat the process in Section \ref{subsec:ivB}, but injecting a reflection-symmetric
% \textit{circularly} polarized 
``GW190521-like'' signal at various SNRs instead. In other words, we inject a sum of damped sinusoids whose encoded polarization structure is constrained by the inclination of a non-precessing source. We aim to demonstrate that the aligned-spin ringdown model can always recover the injected parameters of such signals that arise from truly (anti-)aligned systems and that there is benefit to using this nested model over the simpler, broader generic model. This analysis also serves as a control for the results in the previous subsection, demonstrating that the level of agreement between the two polarization models about the properties of a signal tells us about the symmetry of its source.

To construct the set of 
% circularly polarized injections
aligned-spin model validation injections, we follow a similar procedure to that of Section \ref{subsec:ivB} with minor changes: 1) we construct the $s(t)$ from parameter values inferred by an \textit{aligned-spin} ringdown model fit to the ringdown of GW190521, and 2) the draw we choose from this fit corresponds to the median log-likelihood sample. Thus, by construction, the injection's polarization content in each mode is determined by common $\iota$ and $\Delta \psi$ parameters.  We fit the injections using both the generic and aligned-spin ringdown models; we expect that the results of the two models will be in agreement, but it is interesting to quantify any SNR threshold at which the aligned-spin model may be preferred over the generic model by the data.

We demonstrate in the left panel of Fig.~\ref{fig:injAli_comp} that both polarization models are able to recover the injected $M_f$ and $\chi_f$ across SNRs, and in agreement with one another. Both models produce posteriors that shrink proportionally to the increase in SNR, narrowing around the true values.

The center panel of Fig.~\ref{fig:injAli_comp} compares the inferred to injected amplitude ratios from both models, which agree at all SNRs. The posteriors from both models have long tails to higher amplitudes for both modes at low $B$ due to the modes being out of phase, but these tails are more pronounced in the aligned-spin model results. This may seem at odds with the expected tighter constraints on aligned-spin posteriors given the reduced degrees of freedom in the model. However, the longer tails are artifacts of marginalization over poorly constrained and degenerate parameters like $\Delta\psi$. The parameter $\Delta\psi$ defines the orientation of the $+$ and $\times$ polarization states simultaneously for each ringdown mode in the plane transverse to the direction of GW propagation (\cite{Isi2023}; see App. \ref{subsec:app1}); because $\cos\iota$ is inferred to be face-on, the $|m|=2$ mode is purely circular such that its polarization structure is invariant under rotations of the polarization basis vectors by $\Delta\psi$.
%  \mi{it's difficult to make sense of this discussion given that the previous sections did not introduce the concept of the polarization ellipse, or explained how it relates to $\Delta\psi$ --- not that you have to rederive this, but a short summary with pointers to \cite{Isi2023} would be worthwile}. 
In contrast, the generic model has practically no prior support for circularly polarized ringdown modes ($\epsilon = \pm 1$, see the top panel of Fig.~\ref{fig:GW150914EllipCosi}), so $\theta$ is always fairly well constrained for each mode. Additionally, the $|m|=1$ contribution vanishes entirely for face-on or -off systems; in these cases, the angular factors in Eq.~\eqref{eq:ang_facs} equal zero, allowing the aligned-spin model to overestimate the amplitude of the $(2,1,0)$ mode. So, the large-amplitude tails consist of posterior samples for which the source is inferred to be face-on ($\cos\iota=1$). 
% \mi{very nice}

The right panel of Fig.~\ref{fig:injAli_comp} shows the inference of mode ellipticities by the two models (see the previous section for more details on the reweighting of the generic model posteriors). Both polarization models recover the injected parameters, but the aligned-spin model places much tighter constraints on the ellipticities for all $B$.
% Since the priors are factored out of these distributions, the strong constraints near the boundaries by the aligned-spin model are likelihood-dominated and not a result of prior effects.
This result highlights how sensitively the aligned-spin model can measure the polarization content from the ringdown alone.

\begin{figure}
    \centering
    \includegraphics[width=\linewidth]{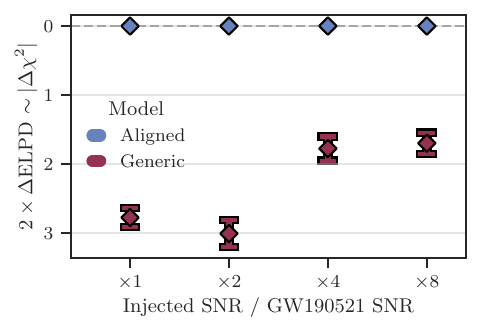}
    \caption{For the set of reflection-symmetric GW190521-like injections, we compare the generic and aligned-spin ringdown models using the data-driven LOO metric. We plot the same difference on the $y$-axis as in Figure \ref{fig:prec_LOO_comp}. The ELPD difference assessment indicates that the aligned-spin ringdown model is preferred by the data at all injected SNRs, or rather the freedoms of the broader generic model are not necessary to describe the data.}
    \label{fig:nonprec_LOO_comp}
\end{figure}

We compare the performance of the two polarization models yet again using the LOO. In Fig.~\ref{fig:nonprec_LOO_comp}, we show that the aligned-spin model is preferred for all $B$ with moderate confidence (1--2$\sigma$, comparable to the strength of the generic model preference for the analyses of the $B \leq 2$ reflection-asymmetric injections). Notably, the values of $\Delta\mathrm{ELPD}$ for each injection are similar to that of the GW150914 results---although the error bars are much smaller, so the evidence for the preferred aligned-spin model is much stronger, as expected for injections into zero noise.
% ; the difference in predictive accuracy between the two polarization models remains roughly constant as a function of SNR, though the strength. 
% The preference for the aligned-spin model is notable in terms of validating its performance \mi{what does ``in terms of validating its performance'' mean?}, but the \textit{weakness} of it suggests that both models are indeed physically valid \mi{phyically valid???} in the case of reflection-symmetric signals.
Increasing the SNR of the injected signal does not increase the preference for the aligned-spin model over the generic model, as was the case for the injections from precessing systems in the previous section.

This asymptotic behavior of $\Delta$ELPD as a function of SNR in Fig.~\ref{fig:nonprec_LOO_comp}, in contrast to the growing difference in Fig.~\ref{fig:prec_LOO_comp}, is well-understood when one model is embedded within another, and the data are consistent with the simpler (aligned-spin) model, as is true here (see Sec.~4 and 5 of \cite{Vehtari+2018}, as well as the left panel of Fig.~10 in \cite{Isi+2017}). Because the predictive distributions are so similar, far more (or louder) data is required to prefer one model strongly based on predictive accuracy than is necessary to precisely measure the parameter differences between the two models. In this case, more informative than the $\Delta$ELPD are the polarization measurements themselves and how harshly those of the generic model depart from the reflection-symmetric constraints of the aligned-spin model. 
% \mi{you may expect a referee to ask if Bayes factors would solve this problem, so you might want to preemptively comment} 
On the other hand, a model comparison using Bayes factors would avoid this asymptotic saturation, since the Occam penalty for the extra degrees of freedom in the generic model would grow with SNR.
While improved detector sensitivity will not make it easier for the LOO to distinguish between the predictions of the two polarization models for data that is consistent with a reflection-symmetric signal, the aligned-spin model will still place tighter constraints on signal polarization content from ringdown inference alone as the instruments become more sensitive over time.
% \mi{this is an artifact of asking a somewhat ill posed question: you are asking the LOO to distinguish between two models that are completely nested, if the two models make essentially the same prediction relative to the data, of course the LOO saturates; that itself tells you you should likely prefer the simple model. In other words, it's just wrong to expect $\Delta$ELPD to keep growing, it's not that we won't benefit from improved sensitivity.}
% \mi{BTW, this well understood and it happens also if you use BFs, e.g., Fig.~10 in \url{https://arxiv.org/pdf/1703.07530}}

We assess the results of the ringdown model comparisons for both the reflection-asymmetric (in Section \ref{subsec:ivB}) and -symmetric injections on the basis of parameter recovery and goodness of fit. Qualitatively evaluating inferred parameters, both models appear consistent with one another between the two sets of injections at the original GW190521-like SNR. However, for signals of SNR around $\times 2$ and louder than that of GW190521, we may be able to identify the aligned-spin ringdown model as invalid for the given signal based on the level of inconsistency between the $90\%$ credible regions of the two models in their inferred remnant and signal parameters. On the other hand, evaluating the model preferred by the data for each injection set, we find that the aligned-spin ringdown model \textit{is} preferred in the case of the reflection-symmetric injections at the GW190521-like SNR, while at the same SNR the generic model is preferred for the reflection-asymmetric injections; at higher SNRs, the generic model becomes overwhelmingly preferred by the data, as we expect once the aligned-spin ringdown model can no longer recover the true values of the parameters. 
% The clear model preference in each case motivates using the LOO as an evaluation metric for the validity of the model.

\begin{figure}[b]
    \includegraphics[width=\linewidth]{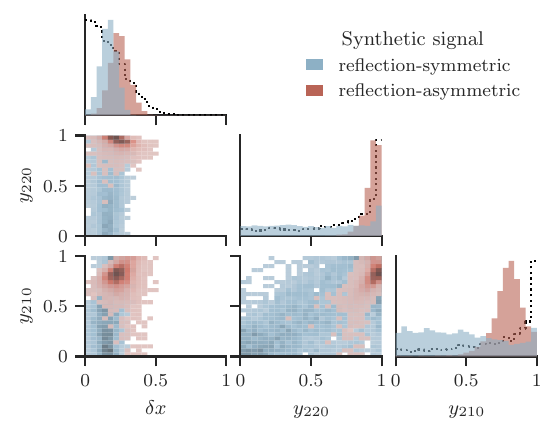}
    \centering
    \caption{Posteriors on parameters that quantify deviations from the aligned-spin parameter subspace, in solid blue and red from fits to the $B=4$ reflection-symmetric and -asymmetric injections, respectively. The priors are shown as dotted black curves in the marginals. See Fig. \ref{fig:GW190521_alisubspace} for more details and conventions. The two sets of posteriors are distinct from one another, demonstrating how sensitive the generic ringdown inference is to the polarization structure of the underlying signal.}
    % The $\delta x$ posterior peaks slightly away from zero. The $y_{210}$ distribution is poorly constrained, while the $y_{220}$ distribution rails against 1 showing very strong support for polarization content that is inaccessible by the aligned-spin model.}
    \label{fig:B4_alisubspace}
\end{figure}

For real GW events we may choose to conservatively use the generic model to perform ringdown analyses, especially if the parameter estimation with IMR templates lacks clear support for a particular progenitor configuration. In this case, we can still leverage the polarization information in the  signals to learn about source properties using the aligned-spin subspace deviation parameters $\delta x_k$ and $y_{\ell |m| n}$. In Fig. \ref{fig:B4_alisubspace}, we show the posteriors on these parameters from generic model fits to the $B=4$ reflection-symmetric (blue) and -asymmetric (red) synthetic GW190521-like signals, as well as the respective priors. Both sets of posteriors are distinct from the priors, as well as from one another. The posteriors from the reflection-symmetric injection fit all have support at zero, the aligned-spin subspace, while the posteriors from the reflection-asymmetric injection fit decidedly rule out the aligned-spin constraints as an adequate description of the signal. We can use the different posterior morphologies for these parameters 
% between two different underlying signals whose polarization structures we tie to their source (a)symmetry, we can use
from such a generic polarization analysis to determine from the ringdown alone whether the remnant is consistent with precessing. For currently achievable SNRs, ringdown inference may not be sufficiently sensitive to the underlying polarization content to \textit{confidently} distinguish between a non-precessing and precessing source using this approach; this is clear from the similarity between the posteriors from the reflection-symmetric $B=4$ injection fit and those of the actual GW190521 data in Fig. \ref{fig:GW190521_alisubspace}. Nevertheless, ringdowns contain accessible polarization information that can be tied to the properties of the progenitor system.

% \mi{it would be nice to check what happens in the space of Fig.~\ref{fig:GW190521_alisubspace} for one of the aligned injections; if that's quite different from what you see in the that fig from GW190521 that's another way of showing that at current SNRs we can already tell that we prefer this non aligned structure.}
\section{Discussion} \label{sec:v}

% \comment[id=MI]{I would start with the most important takeaway which is that you have shown that the ringdown alone contains accessible information about polarization structure from which we can learn about the properties of the source. Then go into more specifics, detailing the model, etc. The aligned model is a tool, which is cool, but the science result comes first.}

In this paper, we have demonstrated that the polarization structure of a GW signal can be sensitively measured from solely the ringdown, and modeling polarization degrees of freedom in ringdown analyses is crucial to extract maximal information about the properties of the source.
We have presented a ringdown model, the aligned-spin model, that can directly and accurately constrain the polarization structure of a GW signal through measurements of the inclination of its source. To do so, the aligned-spin model employs constraints from equatorial reflection symmetry, which parameterize mode ellipticities by a common inclination angle and restrict intrinsic mode amplitudes. The parameter space spanned by the aligned-spin model is embedded within that of the generic (unconstrained) polarization model, in which each mode has four degrees of freedom encoding the amplitudes and phases of the two GW polarizations; on the other hand, the aligned-spin model incorporates fewer degrees of freedom due to the introduction of global polarization parameters that restrict the possible polarization structures of the modes. We assess the aligned-spin analysis of GW150914, which we expect to obey the symmetry of the model, and validate the model's performance for synthetic GW190521-like signals.

% \mi{I would add the punch line in the first sentence: we show that the ringdown alone can be used to infer that the inclination is face-off.}
In the case of GW150914, a system consistent with spin-orbit alignment and equal mass ratio, we show that its ringdown alone can be used to infer that the inclination is face-off. We compare the aligned-spin model's inference of the inclination to that of an IMR analysis using NRSur7dq4. Both analyses infer a face-off inclination, with the ringdown alone requiring that $\cos\iota < 0$ with 97\% credibility. This inclination measurement translates to maximal support for a left-handed circularly polarized signal ($\epsilon_{22n} \approx -1$). 
% The posteriors from the post-peak analysis are broader and do not perfectly overlap with those of the IMR analysis. This slight discrepancy arises from a difference in data and much lower SNR \mi{I don't think this is worth emphasizing here: the model is totally different, of course the posteriors won't exactly match}; 
Despite the ringdown-only analysis fitting a much shorter data segment comprised of a quickly decaying signal with few cycles that are resolvable above instrument noise, the aligned-spin model makes a constrained measurement of the inclination and polarization content of GW150914 without any information from the inspiral and merger.

For this type of signal, the degree of consistency in $\cos\iota$ measurements between the aligned-spin ringdown analysis and the IMR analysis depends on several factors. 
% First, the post-peak data segment used in our analysis is a subset of the full signal that is analyzed with the IMR waveform template; the network matched filter SNR of the reconstructions from the ringdown analysis is ${\sim} 11$, less than half that of the full signal inferred by NRSur7dq4 (${\sim }26$), so it is more difficult to measure the properties of the signal. 
First, the Hanford and Livingston detectors are not ideally arranged to be sensitive to a second GW polarization, rather they are \textit{slightly} misaligned \citep{Riles2013}. Despite this, the IMR analysis is able to infer some polarization structure from information contained in the full signal. The ringdown analysis does not have access to the same polarization information, as it considers only a subset of the same data. The network matched filter SNR of the reconstructions from the ringdown analysis is ${\sim} 11$, less than half that of the full waveform inferred by NRSur7dq4 (${\sim }26$), so it is more difficult to measure the properties of the signal. Furthermore, because various IMR templates model the ringdown under different assumptions about the allowed mode content, they impose potentially incomplete and therefore more rigid relationships between the polarization structure and frequencies in the signal. The ringdown analysis entirely relaxes this assumption such that we can freely identify the signal in the data with any set of modes, subsequently imposing a simple analytic relationship between the ellipticity and frequency observables through a single inclination parameter. This kind of ringdown polarization analysis thus offers another test of the self-consistency of IMR models.
% the aligned-spin model fits a much shorter data segment that contains a quickly-decaying signal comprised of few cycles that are resolvable above instrument noise. With that said, it is a meaningful accomplishment to make a constrained measurement of the inclination and polarization content of GW150914 without any knowledge of the inspiral and merger.
% \mi{this sounds like you're telling yourself this was a meaninguful accomplishment (it is ! good job), but you you should just state the findings, don't editorialize}

For fits to the GW190521-like reflection-asymmetric
% \mi{it's not that the injections are elliptical, you will also get elliptical modes for non face-on/off systems in the aligned-spin model} 
injections, as expected, the aligned-spin model biases measurements of remnant BH mass and spin, mode amplitudes, and ellipticities. While these biased inferred distributions exclude the true injected values of the quantities for SNRs as low as twice that of GW190521,
% \mi{this is meangingless to someone who didn't read the paper, which is most people---say explicitly "SNRs 2x higher than GW190521"}
the data disfavors the aligned-spin model---whose restrictions inaccurately describe the injections' polarization structure (or equivalently, mode amplitude and phase relationships)---for signals as quiet as GW190521 itself.
% \mi{I don't think that's true, the LOO clearly preferred the generic model in all instances, it's just that it didn't make the arbitrary $\Delta = 4$ cut; but definitely looking at the Figures the LOO clearly picks the right model for all SNRs whether the underlying truth is aligned-spin or generic. Is it a 99.7\% confident preference? No, but who cares.}

Understanding the specific failure modes of the aligned-spin model for precessing systems is not necessarily an informative exercise; the discussion in App. \ref{subsec:app4} details how, for a new realization of synthetic injections like those in Sec. \ref{subsec:ivB} but still with similar polarization structure, the aligned-spin model produces qualitatively different biased inferences of remnant and mode properties at different SNR thresholds compared to those in the main text. So, rather than attempting to characterize exactly how the aligned-spin model responds to small variations in the input data in this region of parameter space, we highlight the discrepancies between the aligned-spin and generic models' inference as a robust indicator of a precessing source. 

The quantification of the QNM polarization freedoms' departure from equatorial reflection symmetry constraints serves as a substantive tracer of precession for SNRs greater than that of GW190521; while additional information from the inspiral and merger could provide a stronger measurement, our method offers competitive insight for heavier systems whose GW signals most prominently feature the merger and ringdown. This trade-off can be more carefully studied using injections to obtain forecasts and comparisons of how IMR and ringdown-only precession measurement precision scales with SNR and source properties.

Importantly, the mode content of the synthetic signals studied in this work is fixed a priori, so we do not encounter the usual caveats regarding which QNMs are the appropriate ones to include in our templates. For real data such as GW190521, our polarization analysis method to extract precession information from the ringdown is only as reliable as the mode combination we identify as accurately describing the data. Further validation of that procedure would entail a polarization analysis of synthetic signals constructed by similar means as in Section \ref{sec:iv} but using numerical relativity simulations.

% \comment[id=MI]{I know people address the reader directly like this sometimes, but I hate this---this is not a speech. Just be direct and crisp: ``The generic model should generally serve as the default/baseline since it is always applicable.''}
% We earnestly advise against making restrictive assumptions about the polarization structure of a signal when modeling ringdowns, 
The generic polarization model should generally serve as the baseline template for ringdown analyses as it is always applicable,
unless there is strong evidence for the source exhibiting equatorial reflection symmetry. With that said, there is great benefit to using the aligned-spin model when it is relevant, i.e. in the case of non-precessing systems. For a GW signal from any such system, the aligned-spin model offers a meaningful measurement of the polarization structure, as we have discussed in detail. The model also produces more accurate and precise measurements of remnant and mode parameters, especially if the souce is inclined away from the poles (since this breaks a degeneracy between ellipticity and orientation of the polarization basis). Furthermore, the degrees of freedom in the aligned-spin (generic) model scale with the number of modes included in the fit, $N$, as $2N + 2$ ($4N$); in this work we have only compared our free and restricted polarization models for two-mode fits, but for more modes the parameter space of the generic model grows much faster than that of the aligned-spin model. So, for loud GWs with more modes resolvable above noise in their ringdown signals, the aligned-spin model will be easier to sample and produce increasingly tighter constraints compared to the generic model.

As instrument sensitivity improves and the post-peak SNRs of GW signals increase, the aligned-spin model offers an avenue for a ringdown-only catalog of remnant inference. With this model, we can infer global parameters $\iota$ and $\Delta \psi$ that are directly analogous to those of IMR analyses (unlike the generic model that measures polarization angles for individual modes); it will be especially compelling to compare parameter estimates between our simple analytic template and phenomenological waveform models with more ambiguous prescriptions for the merger-ringdown portion of their templates.

\section*{Acknowledgements} \label{sec:acknowledgements}
We thank Eliot Finch for serving as a reviewer of this work and providing insightful feedback during the internal LVK review process.

This material is based upon work supported by NSF's LIGO Laboratory which is a major facility fully funded by the National Science Foundation.
This research has made use of data or software obtained from the Gravitational Wave Open Science Center (gwosc.org), a service of the LIGO Scientific Collaboration, the Virgo Collaboration, and KAGRA. This material is based upon work supported by NSF's LIGO Laboratory which is a major facility fully funded by the National Science Foundation, as well as the Science and Technology Facilities Council (STFC) of the United Kingdom, the Max-Planck-Society (MPS), and the State of Niedersachsen/Germany for support of the construction of Advanced LIGO and construction and operation of the GEO600 detector. Additional support for Advanced LIGO was provided by the Australian Research Council. Virgo is funded, through the European Gravitational Observatory (EGO), by the French Centre National de Recherche Scientifique (CNRS), the Italian Istituto Nazionale di Fisica Nucleare (INFN) and the Dutch Nikhef, with contributions by institutions from Belgium, Germany, Greece, Hungary, Ireland, Japan, Monaco, Poland, Portugal, Spain. KAGRA is supported by Ministry of Education, Culture, Sports, Science and Technology (MEXT), Japan Society for the Promotion of Science (JSPS) in Japan; National Research Foundation (NRF) and Ministry of Science and ICT (MSIT) in Korea; Academia Sinica (AS) and National Science and Technology Council (NSTC) in Taiwan. The Flatiron Institute is a division of the Simons Foundation.

\bibliography{main}

\appendix
\section{Implementation} \label{subsec:app1}
\subsection{Design matrix}

Our ringdown models sample the real and imaginary parts of complex mode amplitudes in quadratures, or Cartesian components. We can write the plus and cross polarization states for each ringdown mode in the linear basis as
\begin{equation}
    h^{(+/\times)}_j (t) = A^{(+/\times)}_j \cos(2\pi f_j t - \phi^{(+/\times)}_j)\, e^{-t/\tau_j} \, ,
    \label{eq:linpols}
\end{equation}
where $j \equiv (\ell, |m|, n)$. Expanding the difference of angles in the cosine terms in Eq. \ref{eq:linpols} and linearly superposing modes, we can construct the signal template at a given detector (suppressing exponential decay for brevity):
\begin{multline}
    h_{j, \text{ifo}}(t) = \sum_{+/\times} F^{(+/\times)}_{\text{ifo}} A^{(+/\times)}_j \Bigl[ \cos(2\pi f_j t)\cos\phi^{(+/\times)}_j \\+ \,\,\, \sin(2\pi f_j t)\sin\phi^{(+/\times)}_j \Bigr] \, ,
    \label{eq:sig_temp}
\end{multline}
where $F^{(+/\times)}$ are the antenna patterns. We will need to evaluate our likelihood function using this template at each point in the data time series; we can optimize this computation by expressing the template in Eq. \ref{eq:sig_temp} over all observations, modes, and interferometers as a matrix. 

We factor out the four quadratures that enter the template linearly, $x^{(+/\times)} = A^{(+/\times)}\cos\phi^{(+/\times)}$ and $y^{(+/\times)} = A^{(+/\times)}\sin\phi^{(+/\times)}$, and formulate the remaining terms as a design matrix $\mathbf{M}$ with components $M_{isk}$. Enumerating the quadratures with index $q$, the dimensions of the matrix are 
\begin{gather*}
    i \in \{1,...,D\} \\ 
    s \in \{1,...,S\} \\ 
    k = N(q-1)+n, 
    % \\ q \in \{1, 2, 3, 4\}, j \in \{1,...,N\}
\end{gather*}
where $D$ is the number of interferometers, $S$ is the number of observations, and $n$ counts the $N$ modes in the model. So, choosing some $i$, $s$, and $N=1$, the template is given by $h_{\mathrm{ifo}} = (x^+, y^+, x^{\times}, y^{\times})_kM_k$, with
\begin{equation}
    \mathbf{M} =
    \begin{pmatrix}
        F^+\cos(2\pi ft) \\ 
        F^+\sin(2\pi ft) \\ 
        F^{\times}\cos(2\pi ft) \\ 
        F^{\times}\sin(2\pi ft)
    \end{pmatrix}
    \label{eq:DM_n}
\end{equation}
with the rows representing the basis functions for the mode and $t$ being the time of the $s$-th observation.

For the aligned-spin model formalism in the presence of non-precessing symmetry, we can derive in Eq. \ref{eq:ali_pols} the polarizations expressed in terms of common amplitudes and phases. Ultimately, this constraint will reduce the number of quadratures per mode from 4 to 2---one amplitude and phase for both $h_j^+$ and $h_j^\times$. Once again expanding the cosine and sine and grouping terms by quadrature, we can write down the aligned-spin model signal template,
\begin{multline}
    h^{\mathrm{ali}}_{j, \text{ifo}}(t) = \\ \Bigl[F^+_{\mathrm{ifo}} Y^+_j \cos(2\pi f_j t) + F^\times_{\mathrm{ifo}} Y^\times_j \sin(2\pi f_j t) \Bigr] \mathcal{A}_j \cos\phi_j \\ + \Bigl[F^+_{\mathrm{ifo}} Y^+_j \sin(2\pi f_j t) - F^\times_{\mathrm{ifo}} Y^\times_j \cos(2\pi f_j t) \Bigr] \mathcal{A}_j \sin\phi_j \, ,
    \label{eq:sig_temp_ali}
\end{multline}
with the real part of the  intrinsic mode amplitude $\mathcal{A}_j \equiv \bigl|C_{\ell|m|n}\bigr|$, such that our new quadratures are $x = \mathcal{A}\cos\phi$ and $y=\mathcal{A}\sin\phi$ and $q\in{1,2}$. Comparing the expansion in Eq. \ref{eq:sig_temp_ali} to the rows in \ref{eq:DM_n}, it becomes clear how to linearly combine the $k$ element basis functions to construct a new design matrix for the aligned-spin model (for a given $i$, $s$, $N=1$, and $q \in {1,2}$):
\begin{equation}
    \mathbf{M}^{\mathrm{ali}} = 
    \begin{pmatrix}
        F^+ Y^+ \cos(2\pi f t) + F^\times Y^\times \sin(2\pi f t) \\ 
        F^+ Y^+ \sin(2\pi f t) - F^\times Y^\times \cos(2\pi f t)
    \end{pmatrix}
    \, ,
    \label{eq:DM_n_ali}
\end{equation}
such that $h^{\mathrm{ali}}_{\mathrm{ifo}} = (x, y)_kM^{\mathrm{ali}}_k$.

\subsection{Polarization angle}

The global polarization angle $\psi$ positions the ($+/\times$) polarization basis on the sky in the plane perpendicular to the line of slight to the GW source (the wave frame) and determines $F^{(+/\times)}$. In other words, for generic elliptically polarized GWs whose ringdown modes are composed of $\pm m$ QNM contributions with unconstrained amplitudes, $\psi$ fixes the orientation of the $h^{\times}-h^+$ plane in which each mode's polarization ellipse is defined. Any redefinition of the polarization angle by $\Delta\psi$---which manifests as a rotation of the wave frame polarization states by $2\Delta\psi$---can therefore be equivalently absorbed by the angle $\theta_j$ between the semimajor axis of the polarization ellipse and the $h^+$ axis (abscissa) for each mode (see Sec. IVB of \cite{Isi2023} for more detail). For this reason, in the generic ringdown model we fix $\psi$ to an arbitrary value and sample over $\theta_j$ instead \cite{2021arXiv210705609I}. 

In the case of non-precessing systems, the inclination of the orbital plane from the line of sight is fixed in time; this means each mode's polarization ellipse is fixed at a constant orientation in the wave frame, or $\dot{\theta}_j = 0$, and is physically uninformative beyond a redefinition of the polarization basis vectors. So, in the aligned-spin ringdown model, we fix $\theta_j = 0$ (to conveniently align the principal axes of each polarization ellipse with the polarization basis vectors) and restore the polarization degree of freedom $\Delta\psi$. This additional global rotation of the wave frame transforms $F^{(+/\times)}$ according to 
\begin{equation}
    \begin{pmatrix}
        F^+ \\ 
        F^\times
    \end{pmatrix}
    \rightarrow
    \begin{pmatrix}
    \cos(2\Delta\psi) & -\sin(2\Delta\psi)\\ 
        \sin(2\Delta\psi) & \cos(2\Delta\psi)
    \end{pmatrix}
    \begin{pmatrix}
        F^+ \\ 
        F^\times
    \end{pmatrix}
    \, .
    \label{eq:aps_rot}
\end{equation}

\begin{figure}
    \includegraphics[width=\linewidth]{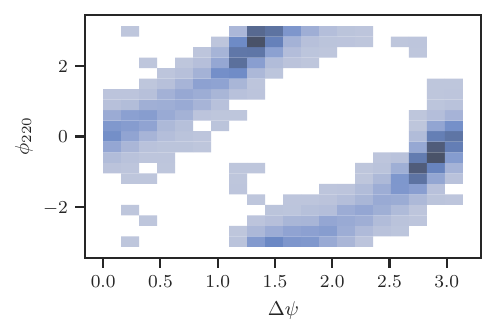}
    \centering
    \caption{Inferred joint polarization angle and 220 intrinsic phase distribution from the Kerr \{220+221\} fit to GW150914 shown in Fig. \ref{fig:GW150914MChi}. The joint posterior is bimodal due to the degeneracy between $(\Delta\psi, \phi_j) \rightarrow (\Delta\psi + \frac{\pi}{2}, \phi_j + \pi)$.}
    \label{fig:pol_phase_degen}
\end{figure}

Note that the transformation depends on $2\Delta\psi$, so the template in Eq. \ref{eq:sig_temp_ali} is unique for all values of $\Delta\psi$ modulo $\pi$. However, recall also the factors of $\cos\phi_j$ and $\sin\phi_j$ in Eq.~\eqref{eq:sig_temp_ali} that scale the transformed antenna patterns; transformations $\phi_j \rightarrow \phi_j + \pi$ introduce a sign change to each term. A redefinition $\Delta\psi \rightarrow \Delta\psi + \pi/2$ flips the signs back, leaving the template unchanged. Fig. \ref{fig:pol_phase_degen} demonstrates this degeneracy in the result of the aligned-spin model fit to GW150914.

\section{Polarization model amplitudes} \label{subsec:app2}

We note a slight difference in the construction of the amplitude parameter inferred by each of the generic and aligned-spin ringdown polarizations models. Returning to the strain of a ringdown mode in Equation \ref{eq:qnm_gen_exp}---in the absence of non-precessing symmetry constraints, the angular factors can be absorbed into redefined complex amplitudes $C'_{\ell m n} \equiv C_{\ell m n} \, _{-2} Y_{\ell m}$. We can then factor these new amplitudes as
\begin{equation}
    C'_{\ell \pm m n} = \frac{1}{2}(1 \pm \epsilon_{\ell |m| n})A_{\ell |m| n}e^{i \phi_{\ell \pm m n}} \, ,
    \label{eq:new_complex_amps}
\end{equation}
where $\epsilon_{\ell |m| n}$ is the same mode ellipticity as in Equation \ref{eq:ellip}, and $A_{\ell |m| n}$ is the real mode amplitude (corresponding to each mode's polarization ellipse semimajor axis) as inferred by the generic model. We can then express the inferred amplitude in terms of the redefined complex amplitudes:
\begin{equation}
    A_{\ell |m| n} = \bigl| C'_{\ell m n} \bigr| + \bigl| C'_{\ell -m n} \bigr| \, .
    \label{eq:generic_amp_from_C}
\end{equation} 

In the generic model, the angular factors are in fact embedded in the amplitudes as we demonstrate above, so the model directly infers $A_{\ell |m| n}$ for each mode. On the other hand, the aligned-spin model keeps the angular structure factored separately, thus instead inferring the magnitudes of the intrinsic complex amplitudes, $\mathcal{A}_{\ell|m|n} \equiv |C_{\ell|m|n}|$.

\begin{figure*}[t]
    \centering
    \includegraphics[width=\linewidth]{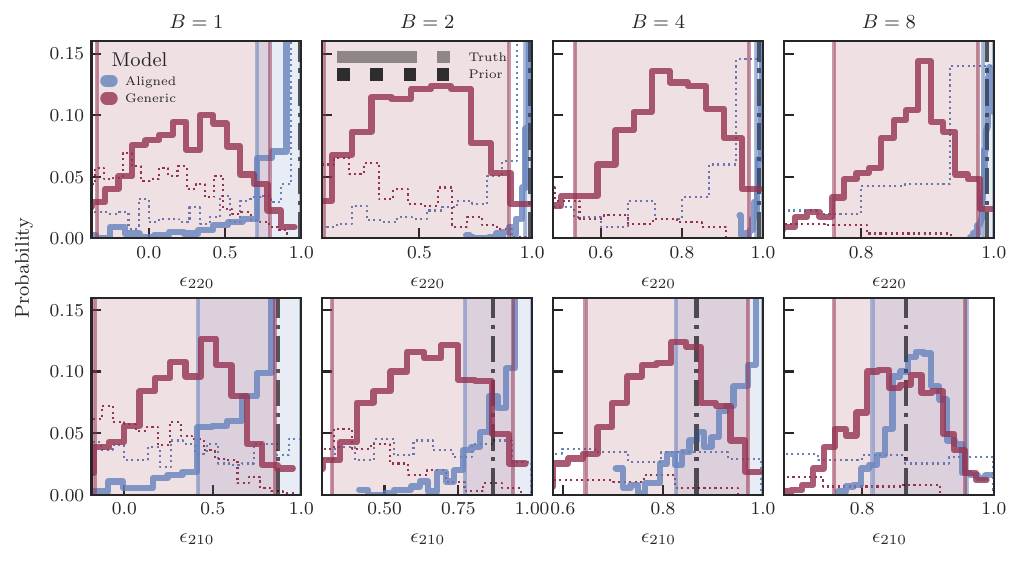}
    \caption{Mode ellipticities inferred by the generic and aligned-spin ringdown models from fits to reflection-symmetric GW190521-like injections. We show the raw $\epsilon_{\ell|m|n}$ posteriors (solid) with the priors (dotted) overlayed. Without reweighting to a uniform prior, the generic model excludes the truth for the $(2, 2, 0)$ at all $B$ and for the $(2,1,0)$ mode at $B=1$ at 90\% credibility due to prior effects, which contradicts our understanding of the aligned-spin model being a subset of the generic model.}
    \label{fig:injAli_ellip_comp_priors}
\end{figure*}
To find the relationship between the two models' mode amplitude parameters, we can solve for $A_{\ell |m| n}$ in Equation \ref{eq:generic_amp_from_C} using the appropriate assumptions imposing equatorial plane symmetry. Expanding both magnitudes, we find
\begin{equation}
    \bigl| C'_{\ell m n} \bigr| = \bigl| C_{\ell m n}\, _{-2}Y_{\ell m}(\iota)\bigr| = \mathcal{A}_{\ell |m| n} \Bigl| \,_{-2}Y_{\ell m}(\iota) \, \Bigr| 
\end{equation}
\begin{multline}
    \bigl| C'_{\ell -m n} \bigr| = \bigl| C_{\ell -m n}\, _{-2}Y_{\ell -m}(\iota)\bigr| = \\ 
    \bigl| (-1)^{\ell} C^{*}_{\ell m n} (-1)^{\ell} \, _{-2}Y^{*}_{\ell m}(\pi - \iota)\bigr| = \\ 
    \mathcal{A}_{\ell |m| n} \bigl| \,_{-2}Y_{\ell m}(\pi - \iota) \, \bigr| \,.
\end{multline}
Here we have used the same parity properties of the spin-weighted spherical harmonics and the complex amplitudes as in Section \ref{sec:ii}. Finally, summing the terms, we get the transformation
\begin{equation}
    A_{\ell m n} = \mathcal{A}_{\ell m n} \Bigl( \bigl| \, _{-2}Y_{\ell m} (\iota) \bigr| + \bigl| \, _{-2}Y_{\ell m} (\pi-\iota) \bigr| \Bigr)
\end{equation}
between the aligned-spin model amplitude parameter $\mathcal{A}$ and the generic model parameter $A$. 

\section{Ellipticity posterior reweighting} \label{subsec:app3}

We note that prior effects may exacerbate inconsistencies in $\epsilon$ posterior comparisons between the generic and aligned-spin ringdown models if the ``true'' value is at the edge of $\epsilon$ parameter space (i.e. $ \pm 1$). For the generic model, the $\epsilon$ prior for each mode goes to zero at the boundaries; while for the aligned-spin model (with a flat prior on $\cos\iota$), the $\epsilon$ prior for $\ell=|m|=2$ modes diverges at the boundaries.  

To demonstrate this effect, we plot in Fig. \ref{fig:injAli_ellip_comp_priors} the $\epsilon_{\ell|m|n}$ marginals (solid histograms) against the priors (dotted histograms) for each model. It is evident that $\pi_{\mathrm{gen}}(\epsilon)$ prevents the generic fits from recovering the injected mode ellipticities (black dash-dotted vertical lines) within 90\% credibility, particularly for the $(2, 2, 0)$ mode. 

Due to this boundary effect in the generic model prior, it is more informative to compare distributions reweighted to a flat prior. To isolate the likelihood contribution and remove prior dependence, we reweight the posteriors by the reciprocal of the prior with weights
\begin{equation}
    w_i(\epsilon) = \frac{1}{\pi_i(\epsilon)} \, , ~ i \in \left[1, \ldots, N_\mathrm{samples} \right]
\end{equation}
such that the resulting distributions inferred by the generic model are proportional to the likelihood functions $\mathcal{L}(d | \epsilon)$. These reweighted distributions are plotted in the right panels of Fig. \ref{fig:injSNR_param_model_comp} and \ref{fig:injAli_comp} in plum.

Note that we cannot reweight to effectively flat ellipticity priors for both modes in the aligned-spin model while preserving the parametrization by $\cos\iota$, since the different $|m|$ of the modes give different transformations according to Eq. \eqref{eq:ellip}. We could instead reweight the generic model posteriors to the by-mode priors on ellipticity induced by the uniform $\cos\iota$ prior of the aligned-spin model; however, the generic model still contains additional degrees of freedom that hinder a direct comparison to the aligned-spin posteriors. 

We proceed with reweighting only the generic model posteriors to uniform ellipticity priors to restore prior support at the domain boundaries. The aligned-spin model inherently has prior support across the entire domain $[-1, 1]$, varying in shape depending on the particular $(\ell, |m|)$ of the mode, as shown by the dotted blue curves in Fig. \ref{fig:injAli_ellip_comp_priors}.
% where $d$ is the data, $s_{\epsilon_{\ell |m|}}$ is the signal, $p(\epsilon_{\ell |m|} \,|\, d)$ is the posterior, and $p(\epsilon_{\ell |m|})$ is the prior. So, in order to compare ellipticity likelihoods, we weight the posteriors by $\frac{1}{p(\epsilon_{\ell |m|})}$ and obtain the results in Figure \ref{fig:injAli_ellip_comp}.

\section{Measure of ``aligned-ness''}\label{subsec:app4}

In Section \ref{sec:iv}, we present the results of aligned-spin model fits to elliptically polarized GW190521-like damped sinusoid injections (we will refer to this set of injections as ``\textbf{inj1}'') of various SNRs, constructed from a posterior sample draw of a generic Kerr $\{(2,2,0),(2,1,0)\}$ fit to the post-peak signal of GW190521. The main takeaway from the analysis is that the aligned-spin model fits to \textbf{inj1} produces biased measurements of all relevant parameters for $B \geq 2$, with the data significantly disfavoring the model for $B \geq 4$. 

To examine how representative this particular posterior sample draw is of the polarization structure of GW190521, we construct a new set of damped sinusoid injections from a different sample draw (we will refer to this set of injections as ``\textbf{inj2}''). We fit the new synthetic signals with the aligned-spin model, following the same procedure as described in Section \ref{sec:iv}. We once again assess the $M_f$, $\chi_f$, $A_{\ell|m|n}$, and $\epsilon_{\ell|m|n}$ posteriors, along with comparisons of the two polarization models' predictive accuracy using the LOO.

Compared to the \textbf{inj1} results, the parameter estimation from the aligned-spin fits to \textbf{inj2} exhibits bias at at different SNR thresholds. The model correctly infers the injected values of $M_f$ and $\chi_f$ for all $B$ in agreement with the generic model (despite longer tails to lower values), biases mode amplitude measurement to slightly lower values for $B \geq 4$, and incorrectly infers mode ellipticities for all $B$. However, the lowest SNR at which the predictive accuracy of the generic model is significantly better than that of the aligned-spin model is $B=2$, which is a higher threshold than that met by the \textbf{inj1} results but the same SNR for which the inferred parameter distributions from fits to \textbf{inj1} become inconsistent between the two polarization models. So, perhaps data-driven model selection is more reliable than parameter estimation comparisons between the generic and aligned-spin models, but only for signals that are roughly 2 times (or more) louder than GW190521.

Because the aligned-spin model parameter estimation ``breaks'' in different ways for the two sets of injections, it is useful to quantify how far they lie outside of the aligned-spin subspace. We will call this measure the ``aligned-ness'' of a particular sample. To quantify it, we must understand the mapping from the generic parameter space to the aligned-spin parameter space. 
\begin{figure}[t]
    \includegraphics[width=\linewidth]{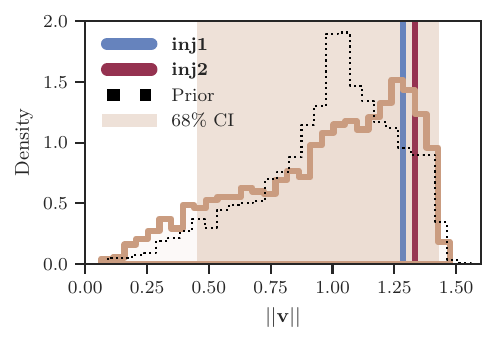}
    \centering
    \caption{We compare the degree of ``aligned-ness'' of each of the posterior sample draws used to construct the GW190521-like injection sets \textbf{inj1} and \textbf{inj2} using the norm of the vector $\vec{v}$ in the space of the equatorial plane symmetry constraint parameters given in Equations \ref{eq:dx} and \ref{eq:y}. $\norm{\vec{v}} = 0$ describes a signal arising from a system that obeys equatorial reflection symmetry, i.e. a non-precessing remnant. So, the greater the $\norm{\vec{v}}$ for a given ringdown waveform, the more it projects into the strictly non-aligned-spin subspace, and the more it violates the constraints of the aligned-spin ringdown model. The tan curve shows the distribution of $\norm{\vec{v}}$ for all posterior samples of the generic ringdown fit to GW190521 from Fig. \ref{fig:GW190521_fit_mchi}. The black dotted line marks the prior, and the blue and plum lines are the sample draws from which we generate \textbf{inj1} and \textbf{inj2}, respectively. The shaded region corresponds to the 68\% credible interval, which contains both sample draws.}
    \label{fig:vnorm}
\end{figure}

From Sec. \ref{sec:ii} and App. \ref{subsec:app1}, we know that the number of free parameters decrease when we map from the generic to the aligned-spin model: the global parameters $\{\iota, \Delta\psi\}$ take the place of the per-mode parameters $\{\epsilon, \theta\}_j$. So for each ringdown mode in the generic model, we can quantify a departure of a particular posterior sample draw from the non-precessing symmetry constraints that enable using the two global parameters for the projection of the signal onto the wave frame. In Sec. \ref{sec:ii}, we introduced two corresponding
% In the generic space, each ringdown mode is parameterized by 4 degrees of freedom: $A_{\ell|m|n}$, $\phi_{\ell|m|n}$, $\epsilon_{\ell|m|n}$, and $\theta_{\ell|m|n}$ \cite{2021arXiv210705609I}. Translating to the aligned-spin space, which spans a subspace of the generic model, $\theta_{\ell|m|n}$ becomes fixed to zero \cite{2023CQGra..40t3001I}, and a new free parameter, the inclination angle $\iota$, parameterizes each $\epsilon_{\ell|m|n}$ via Equation \ref{eq:ellip}. In the aligned-spin subspace, we are left with 3 degrees of freedom: $\mathcal{A}_{\ell|m|n}$, $\phi_{\ell|m|n}$, and $\iota$. For some BBH system, we can assume that the remnant is non-precessing and has some unchanging inclination, permitting us to fix $\iota$. We are left with 2 degrees of freedom, which we can translate into 
``aligned-ness'' parameters---parameters that measure the deviation of a signal's properties from obeying equatorial plane symmetry. To define these parameters, we recalled the constraints of equatorial plane symmetry: 1) mode ellipticities are parameterized by a \textit{common} inclination angle, and 2) intrinsic complex mode amplitudes obey the relationship $C_{\ell-mn} = (-1)^{\ell}C^{*}_{\ell mn}$. 

Using the former constraint, we take the inverse of Eq. \ref{eq:ellip} to convert $\epsilon_{\ell|m|n}$ to $\cos\iota$ and calculate the pairwise differences between modes (similar to the GW150914 $\cos\iota$ posterior comparisons in Sec. \ref{sec:iii}):
\begin{equation}
    \delta x_k = \frac{1}{2}|\Delta\cos\iota|_k\, ,
    \label{eq:dx}
\end{equation}
where $1/2$ is a normalization, and $k$ is the particular pair of modes for which the difference in $\cos\iota$ is computed. If the signal is emitted by a system that obeys equatorial plane symmetry, $\delta x_j = 0$. The complex amplitude constraint gives rise to another set of parameters that we call $y_{\ell|m|n}$:
\begin{equation}
    y_{\ell|m|n} = \frac{\bigl|C_{\ell-|m|n} - (-1)^{\ell}C^{*}_{\ell|m|n}\bigr|}{\bigl|C_{\ell|m|n}\bigr| + \bigl|C_{\ell-|m|n}\bigr|}\, ,
    \label{eq:y}
\end{equation}
where the numerator compares the measured left-handed intrinsic mode amplitude to the left-handed mode amplitude that is required by the non-precessing symmetry amplitude constraint above, and the denominator is a normalization factor. Again, for a ringdown from a non-precessing system, $y_{\ell|m|n} = 0$, and always $y \leq 1$. Note that we can only reconstruct $C'_{\ell m n}$ from the generic polarization template; to compute $y_{\ell|m|n}$ from generic model posteriors, we first obtain the  the intrinsic QNM amplitudes by factoring out $_{-2}Y_{\ell m}(\iota(\epsilon_{\ell |m| n}))$ for which we compute the induced $\iota(\epsilon_{\ell |m| n})$ using the inverse of Eq. \eqref{eq:ellip}.

So, for a given generic ringdown $N$-mode fit posterior sample draw that we choose for constructing the GW190521-like injection sets, we can study its corresponding ``aligned-ness'' vector $\vec{v}$ in the $\frac{1}{2}N(N+1)$-dimensional space that spans all $\delta x_k$ and $y_{\ell|m|n}$ parameters. Namely, we compare $\norm{\vec{v}}$ for \textbf{inj1} and \textbf{inj2} in Fig. \ref{fig:vnorm}. Note there are no posterior samples at $\norm{\vec{v}} = 0$ due to volume effects (the aligned-spin subspace is measure zero in the full space of quadratures), so any sample draw would encode some polarization content that is inaccessible by the aligned-spin ringdown model. 

The $\norm{\vec{v}}$ of the samples used to construct both injections sets are consistent with one another and within $1\sigma$ of the $\norm{\vec{v}}$ distribution. Both samples are representative of the polarization structure of GW190521 as they are drawn from the peak of the $\norm{\vec{v}}$ posterior, which lies at greater values than the peak of the prior.

%%% remove notion of dx from body of the paper %%%

\end{document}